\documentstyle[11pt,myext,rotating,epsfig]{article}

\epsfverbosetrue
\begin{document}           % End of preamble and beginning of text.

\begin{titlepage}
\begin{flushright}
LAPP-EXP 2003.03
\end{flushright}

\vspace{2cm}
\begin{center}
\huge\bf{ A  3-d simulation of the atmospheric neutrinos }\\
\vspace{1cm}
\large                                        
J.Favier $^*$\footnote{$^*${\it\small Corresponding author; e-mail: Jean.Favier@lapp.in2p3.fr}   }
, R.Kossakowski and J.P Vialle\\
\vspace{1cm}

 {\it\small Laboratoire d'Annecy-le-Vieux de Physique des Particules, LAPP,}\\
{\it\small IN2P3-CNRS, BP 110, F-74941 Annecy-le-Vieux, CEDEX, France }

\vspace{1cm}
\end{center}

\begin{abstract}

The first AMS flight in June 1998 on board of the space shuttle Discovery at an
altitude of approximately 380 km
unveiled  unexpected features of the cosmic rays
spectra below the Earth geomagnetic cut-off. In addition to a secondary flux of
particles at all latitude, a ring of high energy particles (up to 6 GeV) and an
anomalous ratio 
e+/e- as high as 4 was observed near the geomagnetic equator. This paper describes a
simulation of the interaction of primary cosmic rays with
atmosphere in which the effect of the Earth magnetic field is included. Using
the GEANT3 package for the tracking of particles with the GFLUKA associated
package for the physics of interactions, this simulation reproduces quite well
the
AMS experimental results and the CAPRICE muon data at ground level. The
predictions of this model for the flux of atmospheric neutrino  are compared with
the Super-Kamiokande results and with the results of other atmospheric neutrino
models.
\end{abstract}
\vspace{1.5cm}
%\end{center}
\end{titlepage}

\newpage
\pagestyle{plain}  

\section{Introduction}
The Super-Kamiokande data on atmospheric neutrinos which show a strong evidence for
$\nu_{\mu}-\nu_{\tau}$ oscillations has stressed the need for accurate
calculation of atmospheric neutrino fluxes and production mechanisms. 
 After pionneering simulation works \cite{honda,gaiss,gaiss2},
 it became obvious that 
the Earth magnetic field has an important effect on the flux
of primary and secondary cosmic rays \cite{batti,batti2,lip3},
, which leads to the need of full 3D simulation.
 Meanwhile, from the measurements of cosmic ray fluxes by the
 AMS\cite{ams1,ams2,ams3,ams4,ams5},
 BESS\cite{bess}, and CAPRICE\cite{caprice1}
experiments, new inputs to the generators had become available, and new
experimental data were also available allowing to check how accurately a model
could predict fluxes. This comprises the measurement of cosmic muon yields at
different latitudes and altitudes by CAPRICE, and since AMS precursor flight on
board of the space shuttle Discovery in 1998, the secondary proton and electron
fluxes coming from the interaction of primary cosmic rays (mainly protons and
alphas) with the Earth atmosphere \cite{ams1}-\cite{ams5}.

Below are presented the results of a full 3D simulation based on the GEANT3.2.1
package, which was used in most of the high energy physics experiments and was
extensively tested in various conditions. In this simulation, primary protons,
alphas, positrons and electrons from outer space in the energy range 0.2 to 500
GeV are traced through the Earth magnetic field. A part of them impinge on the
atmosphere and interact, producing secondary hadrons and leptons. These fluxes
and yields are compared with experimental data. Neutrino distribution
predictions are calculated at the geographic location of the Super-Kamiokande
experiment and compared with the results of other models. The predictions on
absolute fluxes are all normalised from only one parameter, namely the yield of
100 GeV primary cosmic rays measured by AMS\cite{ams5}.

 This work was started in 1999
 to understand the unexpected features of charged cosmic ray fluxes measured by 
 AMS. First
results were presented to the AMS collaboration in November
 1999 reproducing pretty well the AMS data though using only a simple dipole
 to model the Earth magnetic field\cite{favier}.

 \section {Simulation scheme}

 \subsection {Primary generation}
 It is generally admitted that far from Earth, charged cosmic rays density is 
 homogeneous and isotropically distributed in direction, while the rigidity
 distribution is well represented by a power law with a negative exponent close
 to 2.7. The Earth magnetic field
  induces a distortion of this spectrum at rigidity below 30 GV with eventually
  a rigidity cut-off which varies with the geomagnetic latitude. For instance
  to reach the Earth surface at the geomagnetic equator, protons must have a 
  kinetic energy greater than 10 GeV.

  To reproduce by simulation these fluxes without biases, several methods were
  tried. The most straightforward way (so-called method B below), which relies
  only on the properties of cosmic ray distribution mentioned above, is to
  generate particles isotropically on an external sphere of radius R$_{ex}$
  large enough to be beyond the Earth's magnetosphere and to trace the particles
  down to the Earth. Unfortunately, such a method is heavily consuming computer
  time since many particles do not reach the Earth vicinity even if they are
  inside a momentum dependent pre-calculated angular cone (
  fig.\ref{fg:impact}). In the most commonly
  used method (so-called method A below), it is advocated that due to
  Liouville's theorem, particles can be generated isotropically near the Earth's
  surface, despite the spatial distortion induced by the Earth magnetic field on
  momentum and space densities. To take it into account, particles generated
  isotropically near the Earth's surface are charge-inverted and traced back in
  the magnetic field, and only those trajectories which reach the surface of the
  external sphere R$_{ex}$ are kept and then traced forward to Earth, this time
  taking into account interaction processes in the atmosphere. Such a method
  obviously allows to minimize computing time, its efficiency being only governed by
  the cut-off effect.
  
  When the Earth magnetic field and the absorbing Earth are taken into account,
  the validity of method A could be questioned. Thus, before using method A, we
  have made a very high statistic simulation with both A and B methods. The
distribution obtained for key variables were found to be identical (fig.\ref{fg:compa})
, which allowed us to use method A for the rest of the study. It was found that
to avoid distortion in the spectra, the radius R$_{ex}$ of the external sphere
has to be as high as 20 Earth radius ( while most of the authors are using R$_{ex}$
=10 Earth radius for both method A and B, which overestimates by about 5\% the
flux of particle at low energy, as seen on fig.\ref{fg:tkine}).

For the simulation, the Earth is represented by a sphere or radius $R_0$=6371.2
km. The Earth atmosphere is modeled as 82 concentric spherical layers, with a
thickness of 1 km for the 50 first ones and 2 km for the remaining 32
(fig.\ref{fg:calin}).
The air density in each layer is calculated according to the NASA standard model
\cite{atmos} of atmosphere. As a matter of fact, the density of the upper layer
is as low as 0.103 $10^{-7} gr/cm^{3}$. The standard IGRF field\cite{igrf}
representation with Legendre polynomials up to 10$^{th}$ order is used for the
Earth magnetic field, with coefficients corresponding to the date of the AMS01
flight (June 1998). A factor of 30 in tracking computing time was saved by
interpolating this field from a pre-calculated 3D field map grid rather than
calculating the field at each tracking step. We checked that the residuals of
these interpolations are below 0.5\%. The Sun effect on the field is
neglected.
Particles are generated following method A from a spherical layer at 380 km in
altitude, with a radius of the external sphere R$_{ex}$=20 $R_0$

The primary particle spectra (p,$\alpha$, e+,e-) are taken from the measurements
published by the AMS collaboration\cite{ams4}, compatible with a power law
rigidity spectrum of $T^{-2.78}$ at high energy, and the solar modulation is
included as a rigidity-dependent weight\cite{icrc}. To minimize statistical uncertainties
at the generation level, the primaries are generated flat in six energy slices (
in GeV: 0.2-0.5, 0.5-2., 2.-6., 6.-30., 30.-200., 200.-500.), and the events are
weighted during the analysis according to the AMS primary fluxes distributions.
Other components of cosmic rays have been neglected. The accepted primary cosmic
rays representing more than a milion tracks have been recorded on files and used
as input for the simulations of cascades in the atmosphere.

\subsection { Interaction with atmosphere and detection of particles}
The secondary particle fluxes are quite sensitive to the quality of the
description of hadronic interactions. The standard GEISHA hadronic package of
GEANT 3 was found being less accurate than the GFLUKA package when comparing the
predicted secondary particles fluxes to the AMS measured ones for electrons,
positrons and protons. We thus used only GFLUKA for the results shown below.
Neutrinos energy spectra predicted using GFLUKA are also found closer to the
prediction of other models not making use of GEANT 3 for simulation; the ratio
of fluxes obtained with GFLUKA and GEISHA as a function of neutrino energy is
 shown in fig.\ref{fg:geiflu} and reveals deviations up to 30$\%$.

Tracking in the geomagnetic field is stopped when the kinetic energy of the
particle is below the threshold of 100 MeV for gammas and electrons, and 10 MeV
for hadrons and muons. For hadronic interactions with atmosphere, Helium nuclei
are modelled as a bag of 4 independent nucleons each carrying one fourth of the
total energy (commonly known as "superposition scheme"; one of its
justifications can be found in the results of He-p total cross-section at high
energy  which has been  measured
at the CERN ISR\cite{isr} and found to be 4 times the p-p total cross-section.)
 In the interaction of protons or helium nuclei with atmosphere, neutrinos
 produced are assigned the right
 flavour according to their parent particle's nature and tracked forward to the
  Earth surface. Muons are also tracked
 down to the Earth surface for comparison with ground level muon measurements.
 
 For the detection, AMS is represented by a spherical layer at an altitude of
 380 km. Particles are counted as detected if they cross this layer with an
 incidence angle with zenith smaller than 32 deg. for
 protons,
 and 25 deg. for electrons and positrons. This corresponds to the AMS
 experimental angular acceptance. Particle spiralling in the magnetic field are
 counted as many times as they cross the surface within the acceptance angle.
 Statistical error calculation takes into account the fluctuations induced by
 particles with very high weight due to multiple detection.
 
 It is worth mentioning that a global factor of 3 in computing time is saved by using
 method A instead of method B.

\section{Results on charged particles}

The above simulation gives a complete set of prediction on particles fluxes,
either primary or secondary, while using only as input the AMS rigidity
spectra measured at the highest geomagnetic latitude. The dependence with
geomagnetic latitude of the cutoff on energy spectra shows up very clearly
(see fig.\ref{fg:primaire} for primary protons energy spectra at 380 km in
altitude). Secondary particles coming out from interaction of primaries on the
atmosphere produce a second flux below the geomagnetic cut-off. Most of these
interactions occur at an altitude close to 40 km (fig.\ref{fg:alti}), which is a
trade-off between high enough atmosphere density for primary particles to
interact and low enough atmosphere density to avoid absorption of outgoing
secondary particles before their detection at 380 km. For neutrinos which come
from the decay of secondary pions, muons and kaons the probability of
interaction of primaries is dominant and thus the mean altitude of production is
much smaller, about 18 km (fig.\ref{fg:altineut}). Near the geomagnetic equator
the secondary flux of charged particles detected at the AMS altitude is about
doubled. Such excess with respect to higher geomagnetic latitudes is due to the
Earth magnetic field and to its anomalies, in which secondary particles produced
in the atmosphere are brought to higher altitude by following the field lines;
they can oscillate East-West and bounce North-South while spiraling around the
field lines. This creates a kind of ring of charged particle around the altitude
of AMS with an energy as high as 6 GeV. This was observed by AMS for protons,
electrons, positrons, and helium as well.

Fig.\ref{fg:amspro} compares the proton kinetic energy spectra measured by AMS
with the result of this work, as a function of geomagnetic latitude. The
agreement is remarkable enough, though only one normalization constant is used
for all absolutes fluxes, namely the flux of protons at 100 GeV measured by AMS.
Fig.\ref{fg:primaire} exhibits the part of the spectra due to primaries. As
expected, there is almost no particles below the geomagnetic cut-off, which is
not the case for some other works like ref.\cite{vassi} for instance.

Comparison of electron and positron spectra with AMS measurements are shown in
figures \ref{fg:eldown} and \ref{fg:elup}. The agreement is pretty good for
geomagnetic latitudes from 0.3 to 1., with both simulation and real data
showing up the appearance of primary lepton flux above cut-off in downwards
particles. Differently enough, in the bin 0.0 to 0.3 the measured flux of
electrons and positrons is in excess by a factor up to 6 at low energy with
respect to Monte-Carlo prediction. It is worth noting that such discrepancy is
also present in other simulation works \cite{derome}. Although we cannot explain
with certainty this discrepancy, some approximations done in the simulation are
likely to contribute to that. First, the atmosphere density at high altitude
(above 50 km) varies by a factor up to 3 with latitude
(fig\ref{fg:densite}), which can induce a sizeable change in the mean life-time
of the trapped electrons or positrons. Second, it is not possible in the
simulation to trace highly oscillating trapped particles until the end of their
path, which implies an underestimation of their statistical weight. Eventually,
let's remark the difference in absolute flux for upward going and downward
going particles in the experimental data while there is no such difference in
the Monte-Carlo data; if most particles are trapped, upwards and downwards
fluxes should be equal.

The ratio of secondary positron to electron yield, shown in fig.\ref{fg:raelec}
as a function of geomagnetic latitude, is in very good agreement with AMS data.
It exhibits the remarkable feature of varying from about 4 at the geomagnetic
equator down to 1 near the poles. If the Earth magnetic field is switched off in
the simulation, or if the primary flux is restricted to gamma rays, then this
effect disappears. It can thus be deduced that this anomalous ratio
measured in the geomagnetic equatorial region
is due to the effect of the Earth magnetic field on the acceptance of both primary and
secondary particles. Selecting primary protons above 30 GeV gives a ratio of
only 2 at the equator, pointing out the role of the primary proton energy in
this anomaly.

This simulation allowed us to calculate the flux of muons at ground level
and to compare it to the experimental data taken by the CAPRICE
experiment\cite{caprice} in
1994 at Lynn-Lake ( Manitoba, Canada). This is a direct test of the capability
of this simulation to predict neutrino fluxes, since muons and neutrinos are
directly correlated. To take into account the experimental conditions and the
detector acceptance, muons have to be in the geomagnetic latitude range of
64.4-67.3 deg. and to impinge on ground with an angle smaller than 12 deg. with
respect to zenith. To get enough statistics, we integrated on longitude. The
difference of solar activity between June 1998 ( AMS) and July 1994 ( CAPRICE)
as measured by the CLIMAX neutron monitor\cite{climax} (fig.\ref{fg:clim}) is small enough to
neglect its effect on the spectrum of primary protons. Fig.\ref{fg:muon} shows
the comparison between experimental and simulated muon data. The agreement is
pretty good both in shape and in absolute flux, though there is no free
parameter in the simulation.

\section{Atmospheric neutrinos: general features}

The above result on muons is a powerful test of the reliability of the
simulation of neutrinos in our model. As stated above, both the propagation of
primary particles and the development of the hadronic shower have effect on 
neutrino distributions. The magnetic field induces a correlation between
position and direction of charged particles which in turn correlates the various
distributions of neutrinos. For the sake of comparison, many plots shown below
are using the same variables and cuts than those used by other authors.
Distributions are shown in 3 latitudes slices ( in radians: 0.-0.2, 0.2-0.6,
0.6-1.) and in 4 neutrino momentum ranges( in GeV: 0.1-0.31, 0.31-1.,1.-3.,
3.1-10.). We call " zenith angle " the angle between the zenith direction in a
given point and the direction of a neutrino arriving  at this point ( thus, a
neutrino going in direction of the Earth's center has a zenith angle of 180
deg.).

Fig. \ref{fg:altineut} shows the altitude of production of neutrinos from pion
and muon decays. On average, neutrinos from muon decay are produced 2 km lower
in altitude, due to the longer lifetime of muons. Near the geomagnetic poles,
neutrino flux is higher by about a factor 2 at low energy, which is related to
a higher primary proton flux corresponding to  the
lower geomagnetic cut-off present in these regions ( fig.\ref{fg:enlat}). The
predictions on zenith angle of neutrinos are shown in fig.\ref{fg:zenith1} to
fig.\ref{fg:zenith3}. There is a good compatibility in shape with some previous
works \cite{lip3,buen3,honda2}, though there is differences
on amplitudes. This could come for instance from the simulation of hadronic
cascades, but further work is needed to identify precisely the origin of these
differences. An enhancement
of the zenith angle distribution is seen at 90 deg. in all plots. Such effect
was not present in former simulations using a 1 dimension approximation. The
azimuth distributions are shown in fig.\ref{fg:azi1} to fig.\ref{fg:azi3}. In
all these plots, asymmetries fade out when energy increase, reflecting also
the higher energy of the parent primary, less and less sensitive to magnetic
field effects.

Last, fig.\ref{fg:razen1} to \ref{fg:razen3} show the ratio of muonic to
electronic neutrinos ($\nu_{\mu}+{\bar\nu_{\mu}}$)/($\nu_{e}+{\bar\nu_{e}}$) for
the same latitude and energy bins. The strong zenith angle dependence which
shows up is due to the correlation between this angle and the path length of
parents particles. Compared to other works \cite{buen3}, we find a smaller ratio
in the vertical direction ( zenith angle near 0 or 180 deg.). As an example,
this ratio goes only up to 3.5 in our data ( energy slice 3.1-10. GeV, latitude
0.2-0.6 rad.) while \cite{buen3} gets a value of 4.2.

\subsection{Results in the Super-Kamiokande region}

The Super-Kamiokande  region is approximated by a slice in geomagnetic latitude
covering $\pm5$ deg. around the geographic location of the Super-Kamiokande (S.K.)
experiment. The flux is integrated on longitude to get more statistics. We
checked that there is no visible difference when the longitude is restricted to
$\pm$30 deg. around the S.K. one.

In S.K., the neutrino variables are deduced from the measurement of the charged
lepton produced in the interaction with the sensitive medium. The reconstruction
accuracy is pretty poor below 1 GeV. To take the measurement error into account,
the neutrino direction generated is smeared \cite{learned,sobel} according
to the law 40$^o/\sqrt{(E_{\nu}}$ to get the outgoing lepton direction, while
for the lepton momentum a flat random generation between 0.25$\times{E}_{\nu}$
and E$_{\nu}$ is used. The S.K. detection Monte-Carlo would be necessary for
getting a better approximation. To normalize our data to the number of events in
S.K., the simulated events are weighted by the cross-section of neutrinos on the
sensitive medium, which at very low energy is only poorly known\cite{crossec}.
The effect of smearing can be seen in comparing figures \ref{fg:zenit}
and \ref{fg:zenitsk} which show zenith angle distribution for two energy
ranges. For the lower energy bin, peaks are washed out when smearing is applied.

The fig.\ref{fg:hondspec} shows the simulated neutrino fluxes multiplied by 
$E_\nu^{2.5}$. Here we use for the sake of comparison the weights corresponding
to primary spectra used by Honda et al.\cite{honda}; the agreement between our
result and the first results of Honda et al.\cite{honda} and Bartol group\cite{gaiss}
 is very good. In figure \ref{fg:ninospec}, we use AMS primary spectra as we did
 above, with a solar modulation correction corresponding to years 1996,97, and
 98 ( see fig.\ref{fg:climsk}). The results of other works (\cite{honda},\cite{batti},\cite{buen3}) are
 also shown. Contrary to the previous figure, our results in fig.\ref{fg:ninospec} are sensibly higher
 than the other ones for energies below 0.4 GeV, while the agreement is good for
 higher energies. We are in general 20\% higher in flux than the recent results
 of G. Battistoni et al.\cite{batti}. To test the sensitivity of these spectra
 to the solar modulation, we calculated the neutrino energy spectra with the
 parameters corresponding to years 1999 to 2001. The ratio of the former
 predicted fluxes to these new ones is as high as 10\% at low energy, going down
 to 5\% above 1 GeV. As for the rest of this paper, the Honda's parametrization
\cite{honda} of the solar modulation is used.

The ratio of fluxes (up-down)/(up+down) for muonic and electronic neutrinos is
shown in fig.\ref{fg:updown}. Events with zenith angle in the range -0.2 +0.2
are not used. In the top histograms, events generated are not smeared, while
bottom ones are smeared for lepton measurement according to the formulae
described above. Unsmeared muonic neutrino data are in quite good agreement with
the recent work of G. Battistoni\cite{batti2}. The good agreement between our
prediction and S.K. data for electronic neutrino, and the very bad one for
muonic neutrino confirm the evidence for a muonic neutrino disappearance.  Fig. 
\ref{fg:zenitsk} shows the comparison between data and Monte-Carlo for the
zenith angle distribution for the  " sub-GeV" class of events 
and the " multi-GeV" one. Electronic-neutrino distribution is
normalized to equal surface, while for muonic neutrinos the normalization is done
only on the last 3 bins in zenith angle. Our distribution is in slightly better
agreement than the results of previous works. This gives a hint about the
sensitivity of the S.K. oscillation parameter result to the model used for simulation
. However, it could be that this is dominated by the way the smearing is taken
into account.

The ratio of fluxes of neutrino species is given in fig. \ref{fg:rapsk} as a
function of zenith angle. We have not been able to compare it with S.K. data
since this apparatus has not the same acceptance for electrons and muon, the
S.K. detector Monte-Carlo being necessary to evaluate them.
 Our ratio is smaller than in other
works, which could lead to a change in the oscillation
parameters calculated by S.K. with their simulation.

Also shown (fig.\ref{fg:estouest}) is the azimuthal distribution of neutrino
species compared with S.K. data, in quite good agreement for
electron-like type and exhibiting the so-called " East-West effect", and with some
difference for muon-like neutrino events. ( we did not put $\nu$ oscillation in our events to look for a
possible explanation of this disagreement).

\section{Conclusion}

A model of simulation of the effect of interaction of primary cosmic rays with
the Earth atmosphere has been developped. Based on the GEANT3 package together
with its GFLUKA option for hadronic interactions, and using the primary energy
spectra measured in AMS as input, this work was able to reproduce pretty well
all the distribution measured in AMS, while using the proton flux at 100 GeV
measured by AMS as unique normalization constant. The difference on the absolute
flux of leptons predicted at the geomagnetic equator is likely to have its
origin in the variation of atmosphere density with geographic latitude which is
not taken into account in this work. However the effect is sizeable only at low
energy.

This work has allowed also to predict fluxes of neutrinos from interaction of
primary cosmic rays on atmosphere. The good agreement between the prediction on
muon fluxes at ground level and the measurements of the CAPRICE experiment
demonstrates that the neutrino simulation is reliable and accurate. It is
compatible with most of the other simulation work within 20\%. The agreement
with the Super-Kamiokande e-type neutrinos data is quite good, while a clear discrepancy
is observed for muon-neutrinos, as expected from neutrino oscillation. As a
conclusion, the secondary flux of particles observed by AMS does not affect the
conclusion of the Super-Kamiokande experiment about neutrino oscillations.
However, the slight differences observed between the results of this work and
those of the other
authors  
indicate the kind of room left for systematic error on the proposed oscillation
parameters; but this needs the S.K. apparatus simulation to be precisely evaluated.

\section{Aknowledgements}
We would like to thank L. Derome, F. Donato, M. Maire and N. Produit for helpful
 discussions.

%fig 1
\newpage
\begin{fighere}
\begin{center}
 \mbox{\epsfig{file=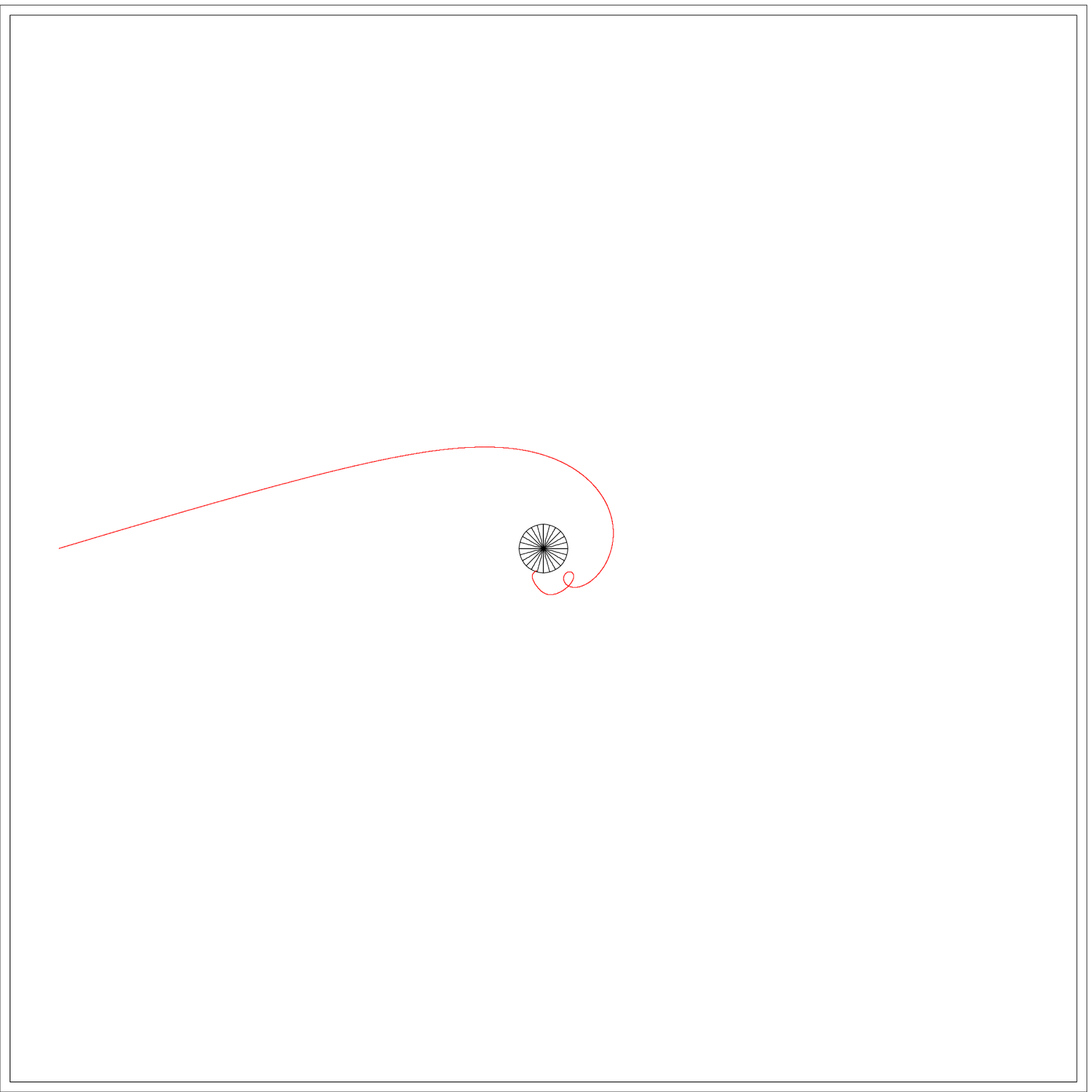,width=16cm}}
\end{center}
\caption{\small Trajectory of a 5 GeV proton generated with an impact parameter
of 6$R_{0}$, but nevertheless reaching Earth. Generation far from Earth (method
B) requires such large aperture generation cones leading to poor efficiency.
}
\label{fg:impact}
\end{fighere}

%fig 2
\newpage
\begin{fighere}
\begin{center}
 \mbox{\epsfig{file=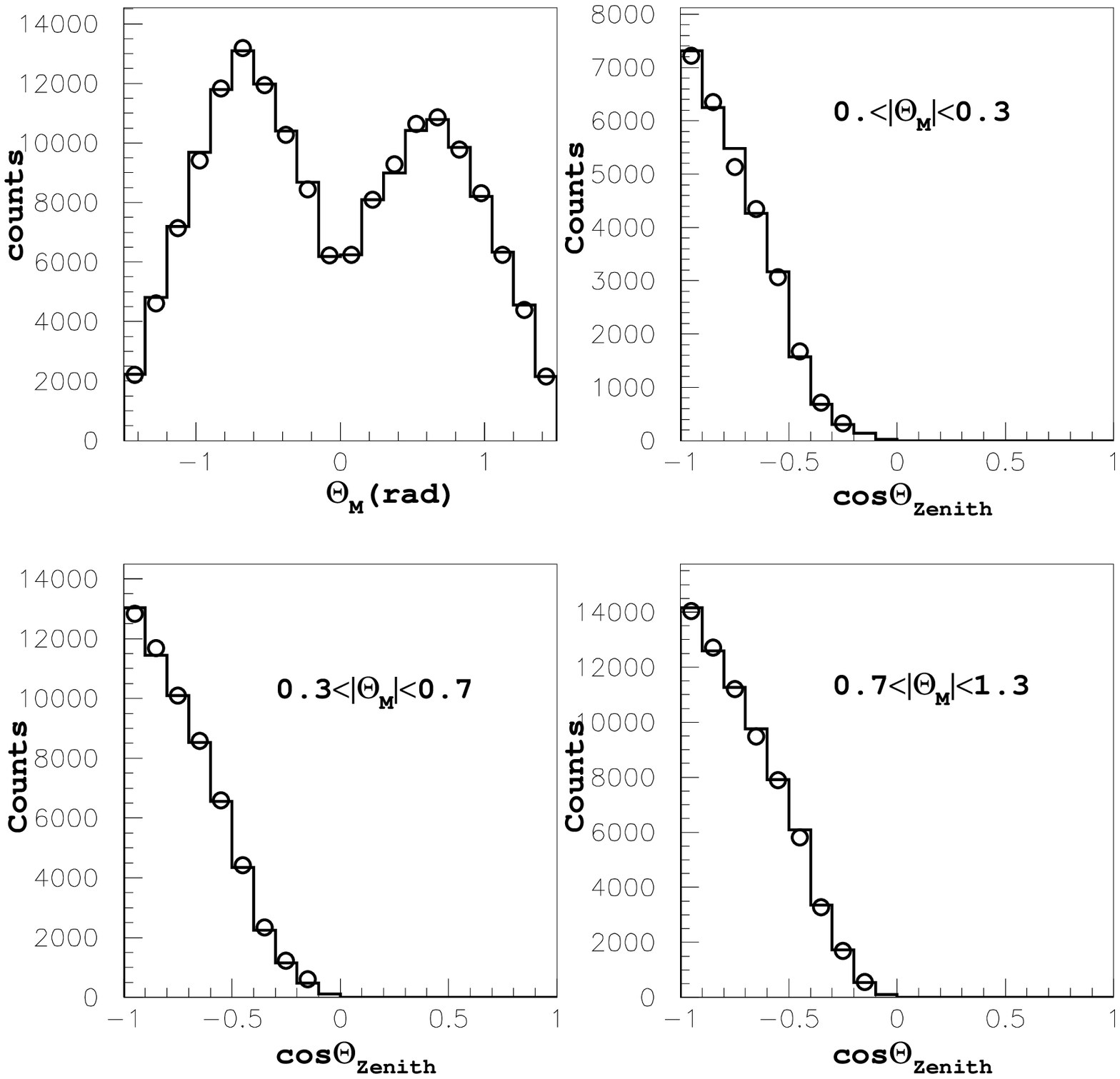,width=16cm}}
\end{center}
\caption{\small Comparison of generation far from Earth  and near Earth ( open
circles); (a) is
the geomagnetic latitude population; the observed distortion of symmetry is a reflection of
the difference of a dipole field and the real one which
contains also the South-Atlantic anomaly. (b), (c), (d) are distributions of the zenith
angle for three geomagnetic latitude bins. Positions and angles are computed at
the 380 km
altitude and here the proton energy range is 6-30 GeV.
}
\label{fg:compa}
\end{fighere}

%fig 3
\newpage
\begin{fighere}
\begin{center}
 \mbox{\epsfig{file=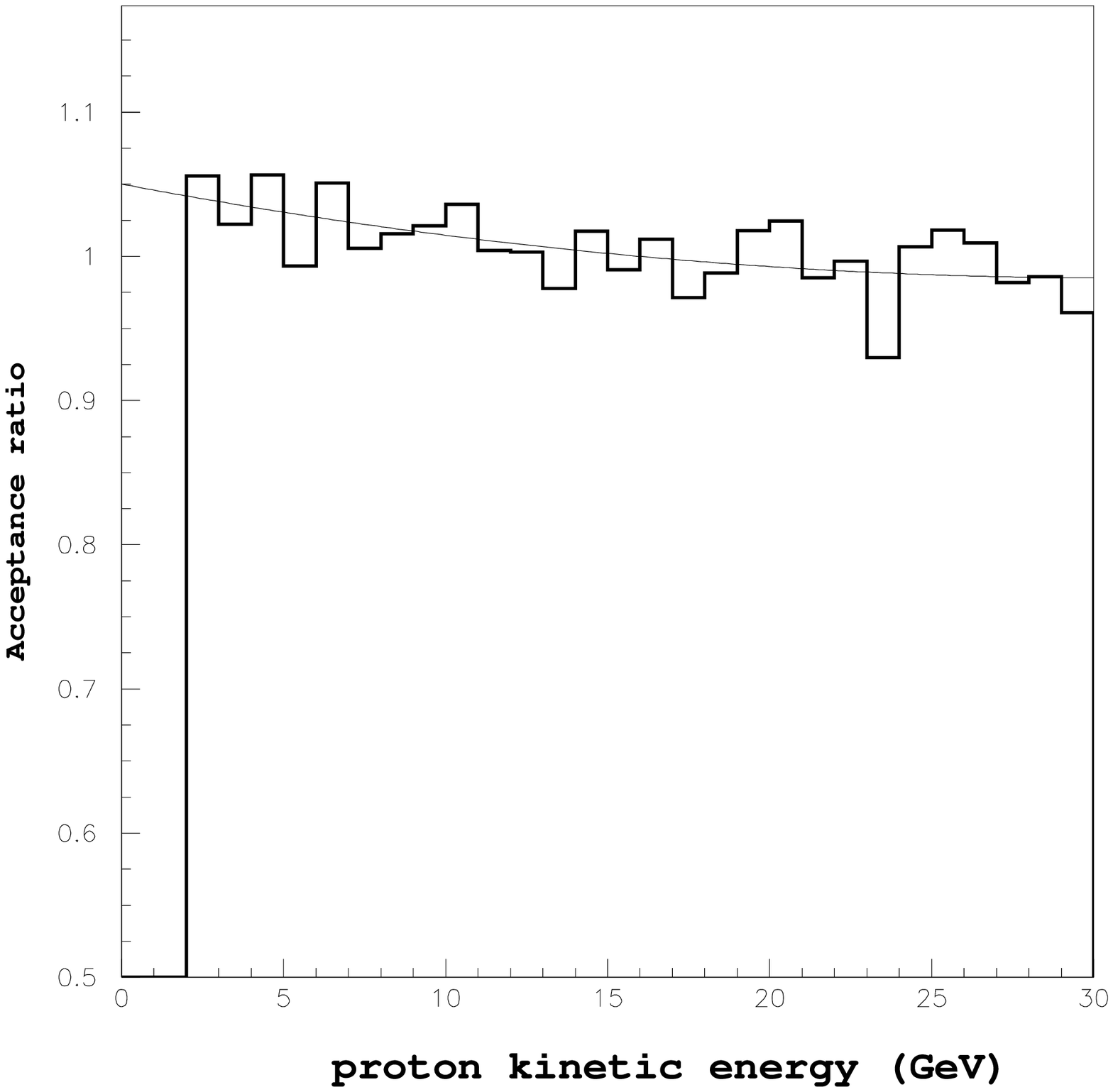,width=16cm}}
\end{center}
\caption{\small Ratio of acceptance between generating primary cosmic rays with
an external sphere of 10 $R_0$ and
20 $R_0$. The higher acceptance for the nearest distance means the presence of wrongly accepted
trajectories.}
\label{fg:tkine}
\end{fighere}

%fig 4
\newpage
\begin{fighere}
\begin{center}
 \mbox{\epsfig{file=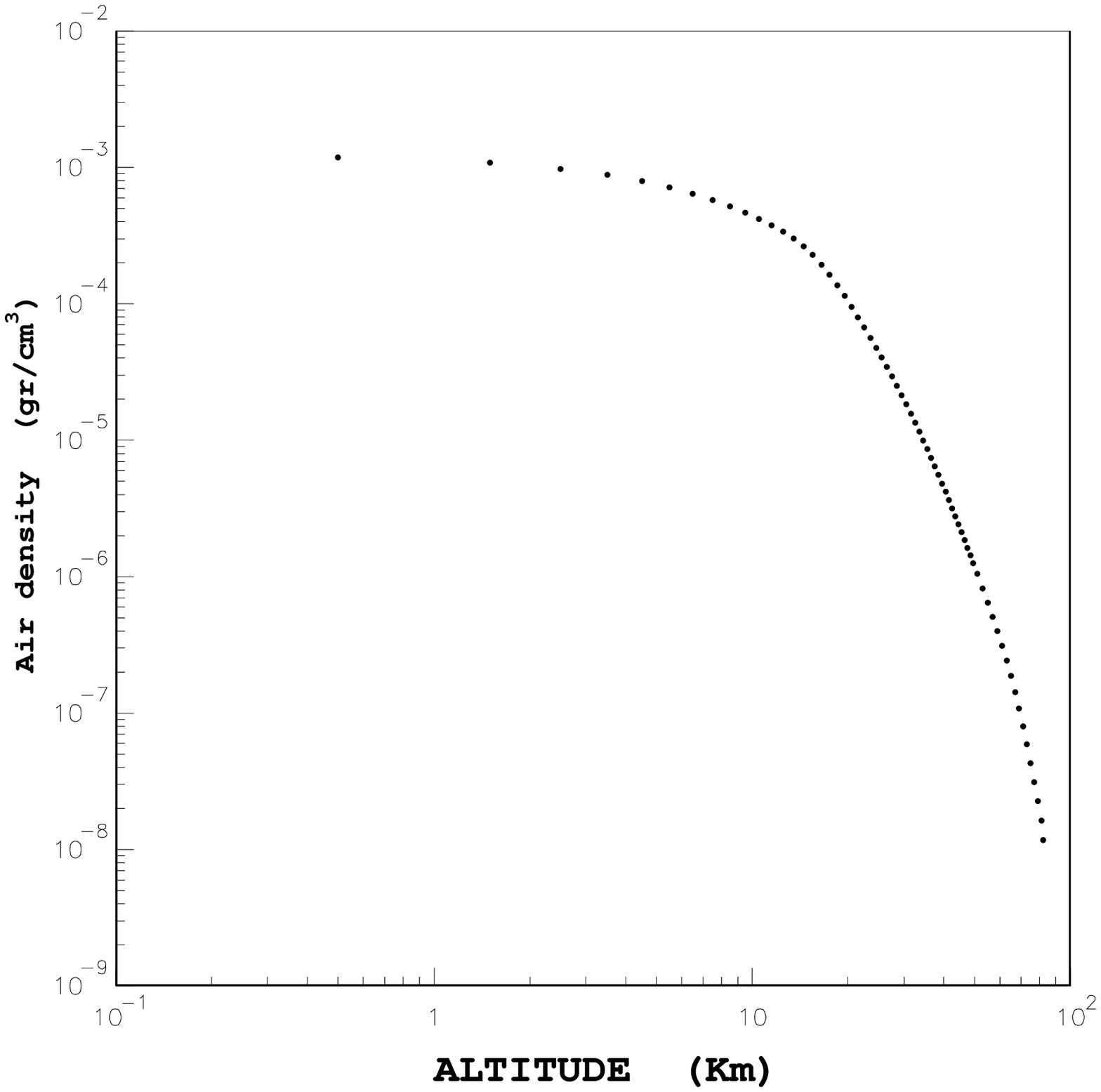,width=16cm}}
\end{center}
\caption{\small Air density as a function of altitude ( data from \cite{atmos})
}
\label{fg:calin}
\end{fighere}

%fig 5
\newpage
\begin{fighere}
\begin{center}
 \mbox{\epsfig{file=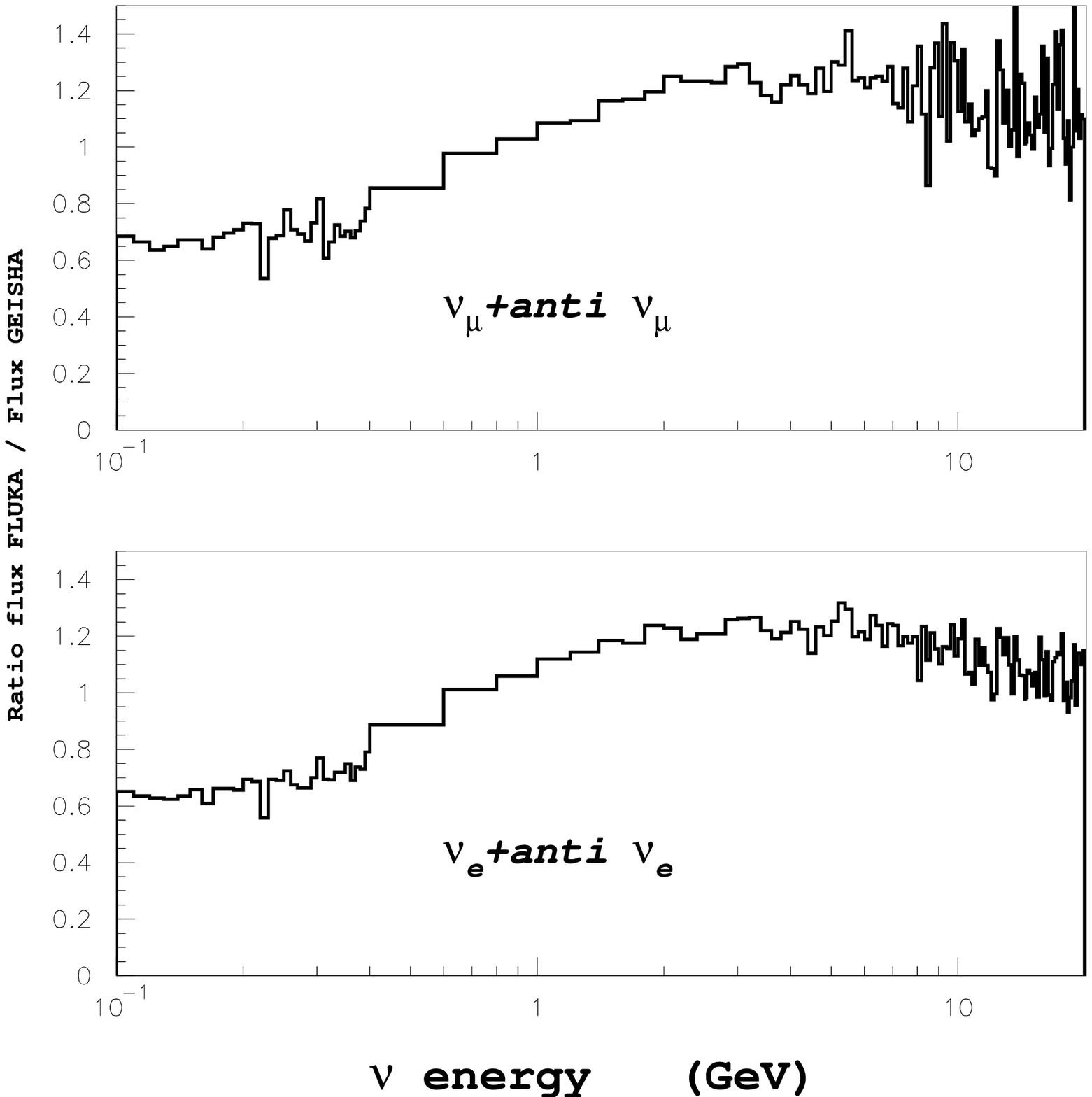,width=16cm}}
\end{center}
\caption{\small Ratio of atmospheric neutrino fluxes for a generation using
GEANT3+GFLUKA package and a generation using GEANT3+GEISHA package as a function
of neutrino energy for all geomagnetic latitudes.
}
\label{fg:geiflu}
\end{fighere}

%fig 6
\newpage
\begin{fighere}
\begin{center}
\mbox{\epsfig{file=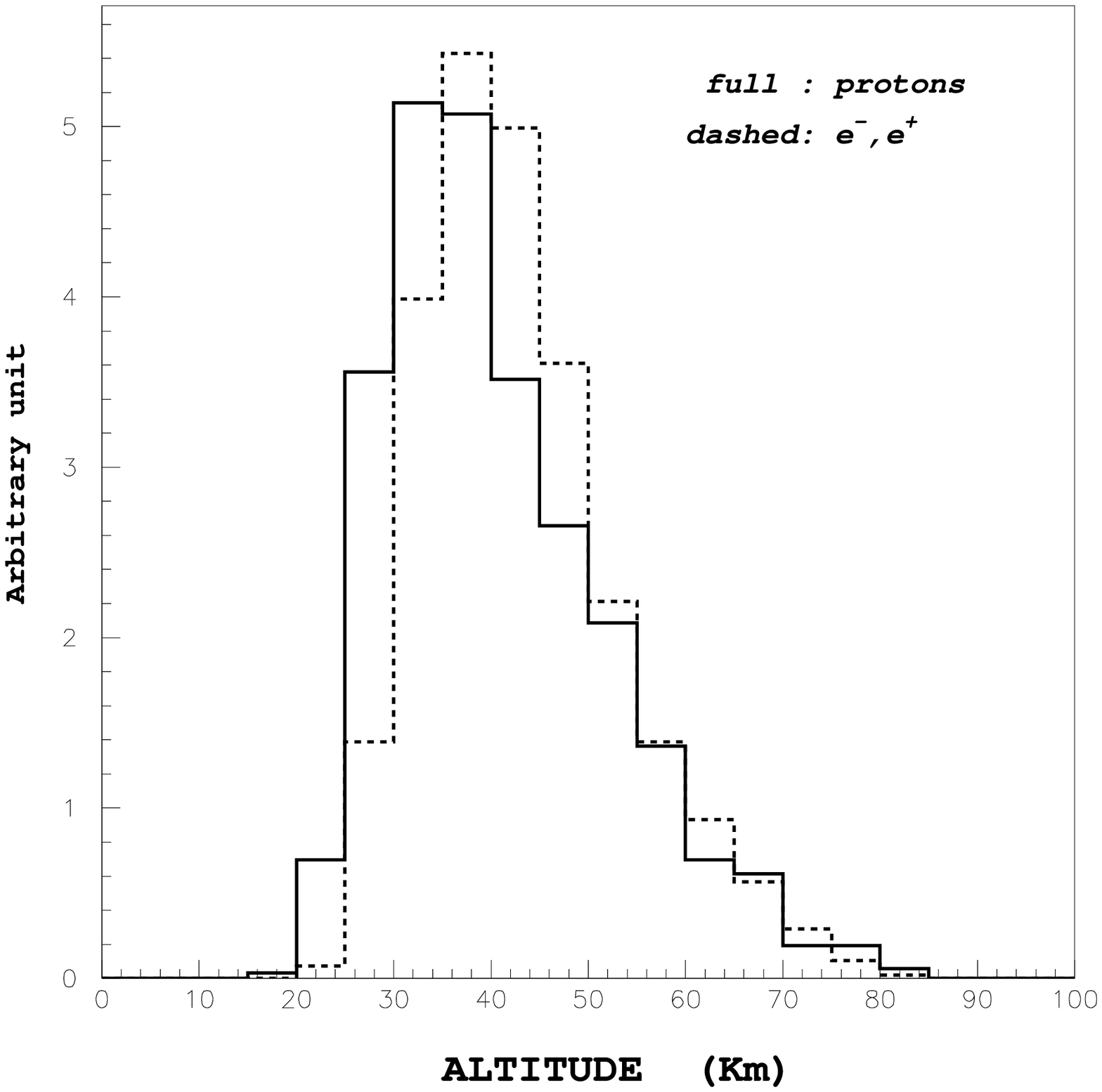,width=16cm}}
\end{center}
\caption{\small  Production vertex altitude distribution for secondary protons
and for electrons or positrons; average altitude is 41 km for secondary protons
and 43 km for $e^-$ and $e^+$.( These plots contain secondaries detected at the
 shuttle altitude with no angular
acceptance cuts)  
}
\label{fg:alti}
\end{fighere}

%

%fig 7
\newpage
\begin{fighere}
\begin{center}
 \mbox{\epsfig{file=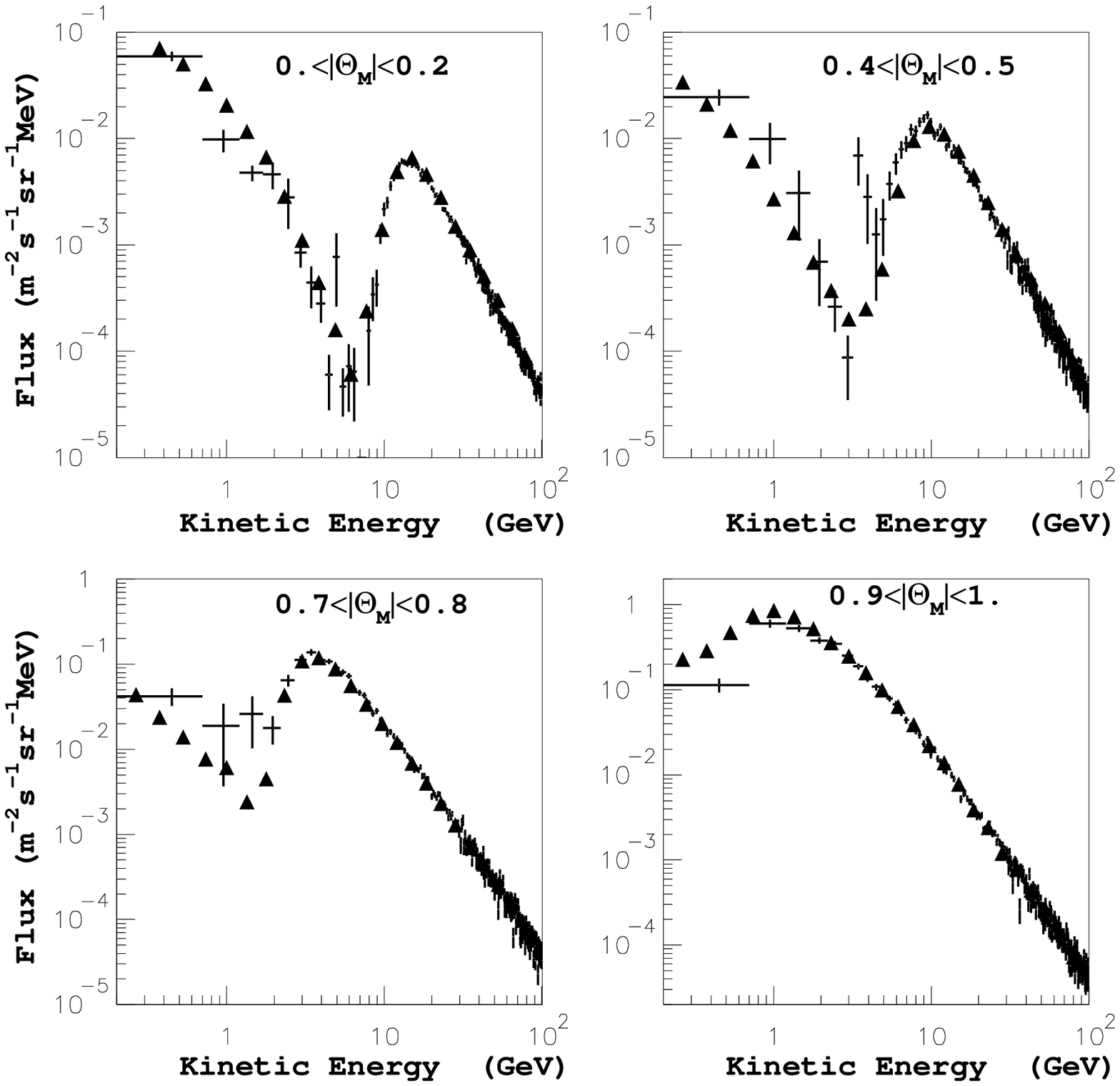,width=16cm}}
\end{center}
\caption{\small Comparison at different latitudes of AMS proton spectrum (black
triangles) and this simulation. The MC spectrum is
normalised once to the AMS data in the 100 GeV energy region. 
}
\label{fg:amspro}
\end{fighere}

%fig 8
\newpage
\begin{fighere}
\begin{center}
 \mbox{\epsfig{file=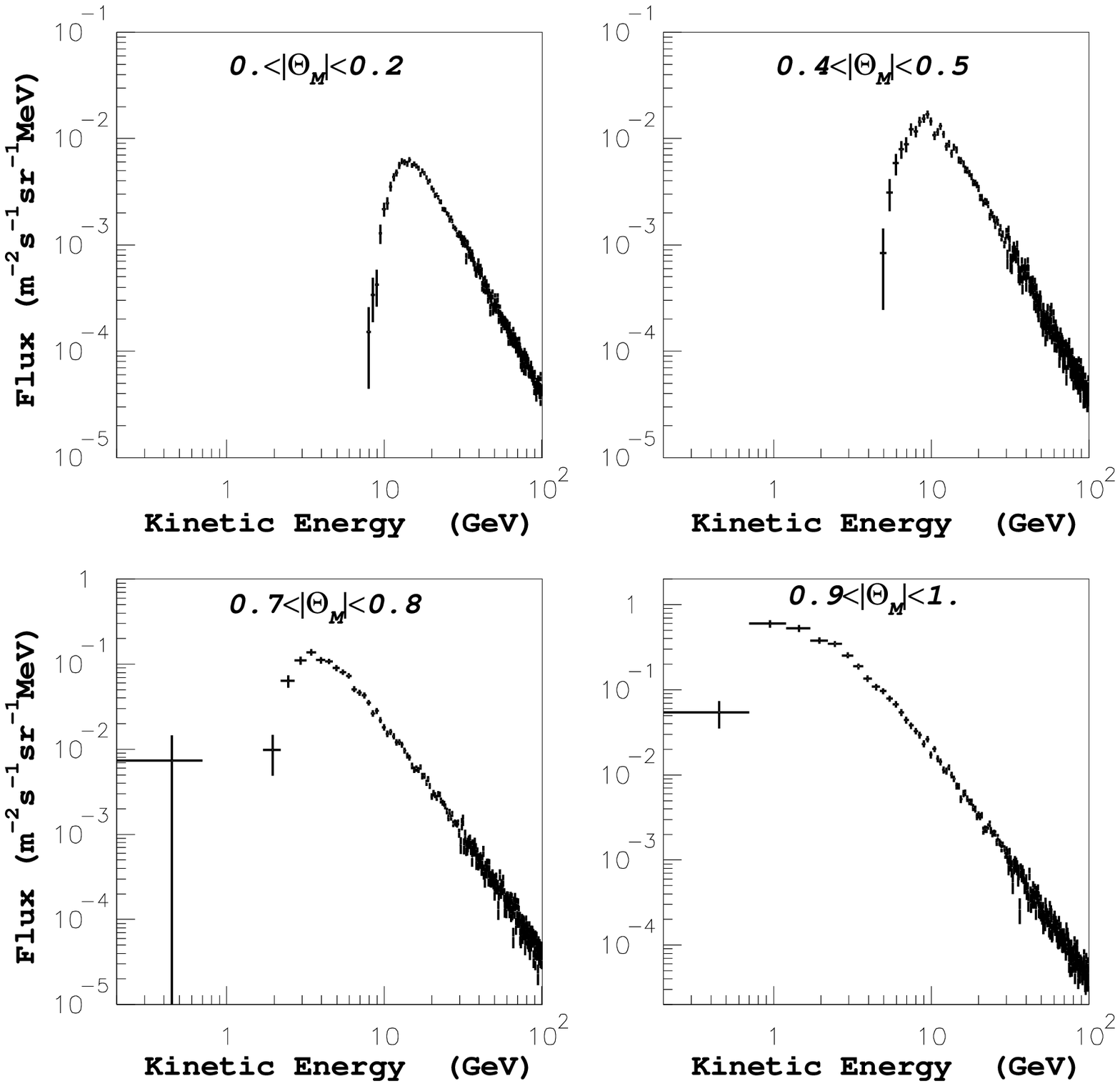,width=16cm}}
\end{center}
\caption{\small Same spectra as figure \ref{fg:amspro} in selecting only primary
protons; the displacement of the geomagnetic cut-off with latitude is clearly
seen. 
}
\label{fg:primaire}
\end{fighere}
%fig 9
\newpage
\begin{fighere}
\begin{center}
 \mbox{\epsfig{file=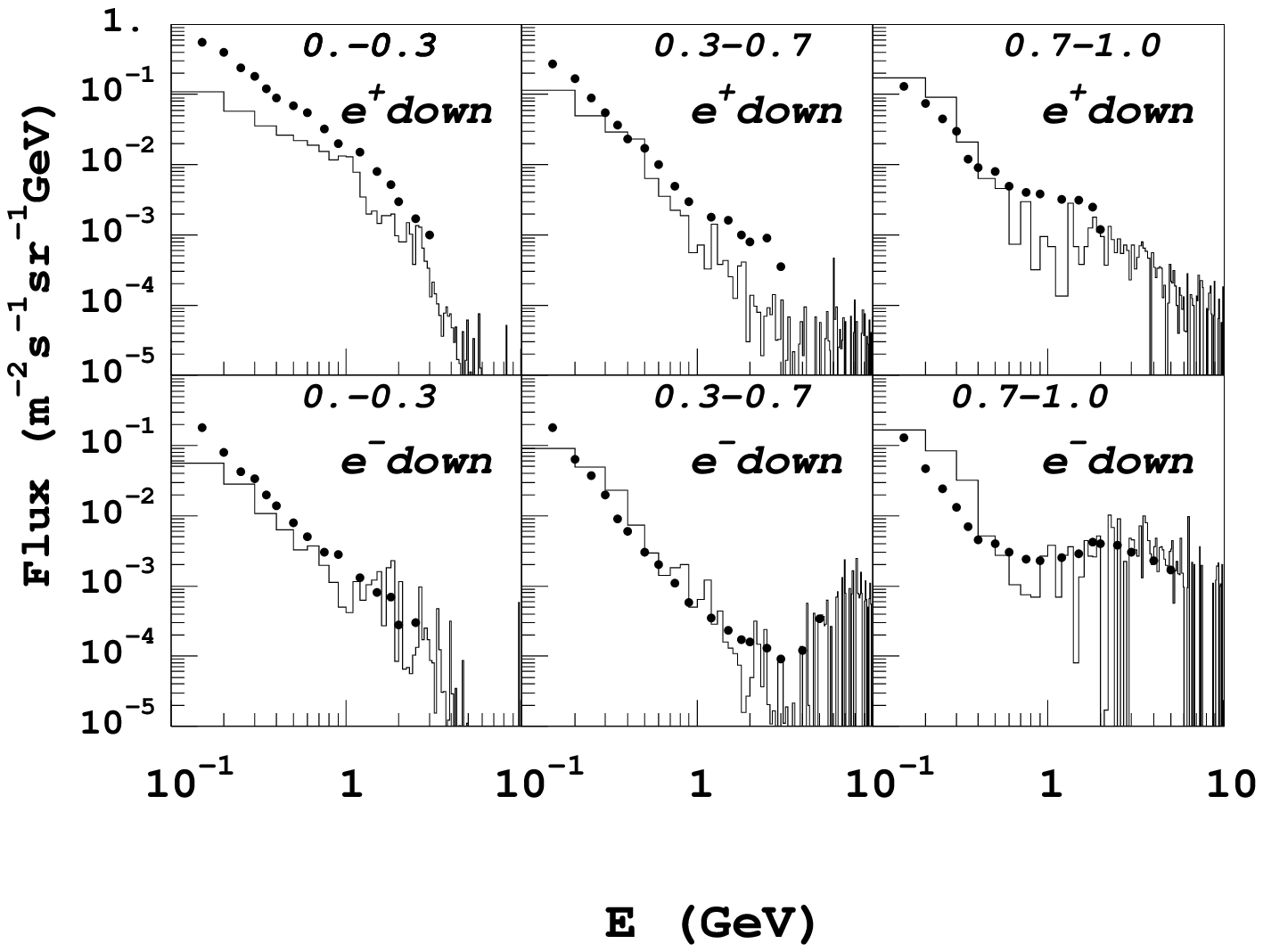,width=16cm}}
\end{center}
\caption{\small Downwards electron and positrons energy spectra for different
geomagnetic latitudes; black dots are
AMS data; notice the appearance of the primary electrons above cut-off. }
\label{fg:eldown}
\end{fighere}

%fig 10
\newpage
\begin{fighere}
\begin{center}
 \mbox{\epsfig{file=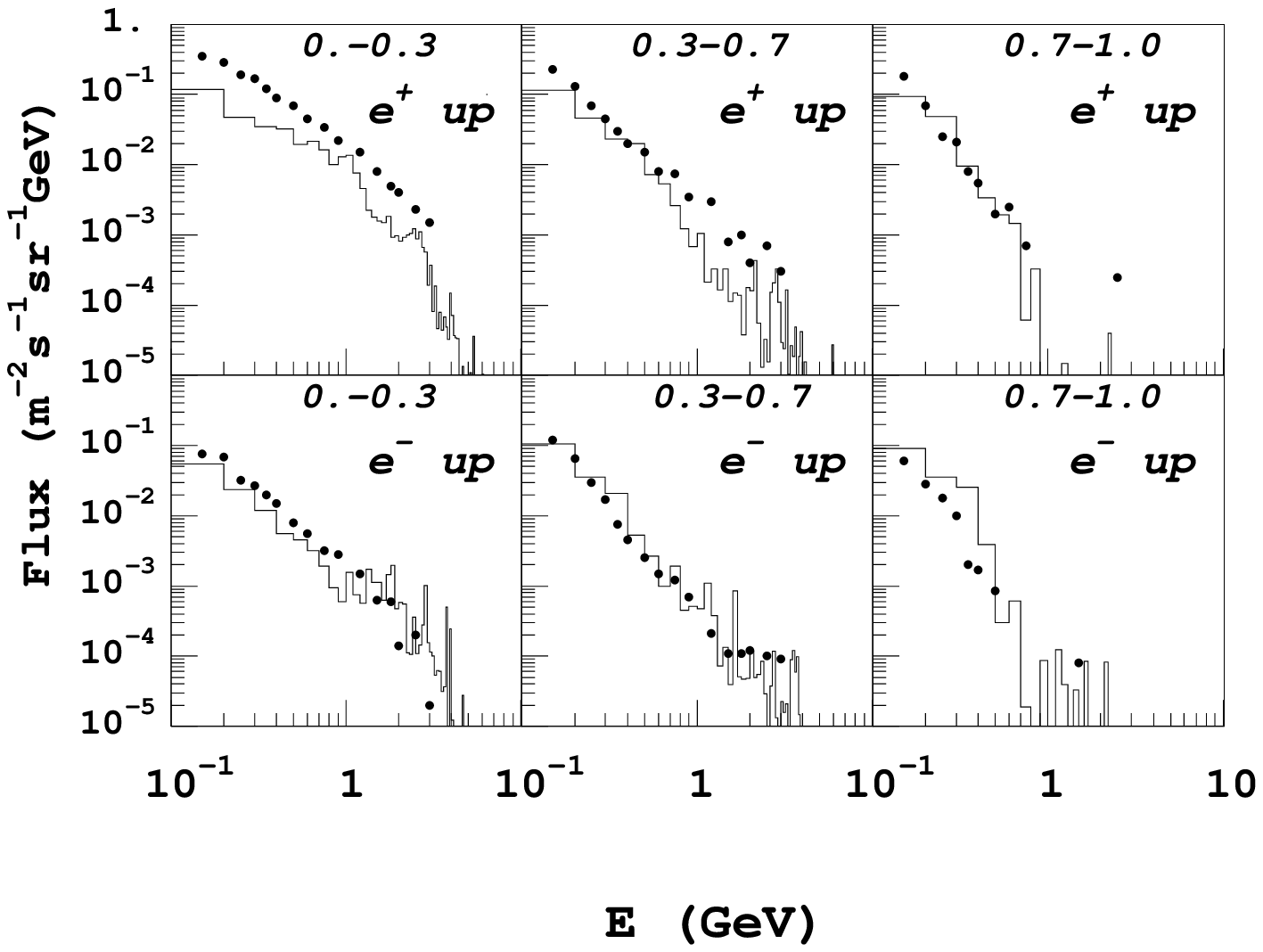,width=16cm}}
\end{center}
\caption{\small Upwards electron and positrons energy spectra for different
geomagnetic latitudes; black dots are AMS data ( in its identification range)
}
\label{fg:elup}
\end{fighere}
%fig 11
\newpage
\begin{fighere}
\begin{center}
 \mbox{\epsfig{file=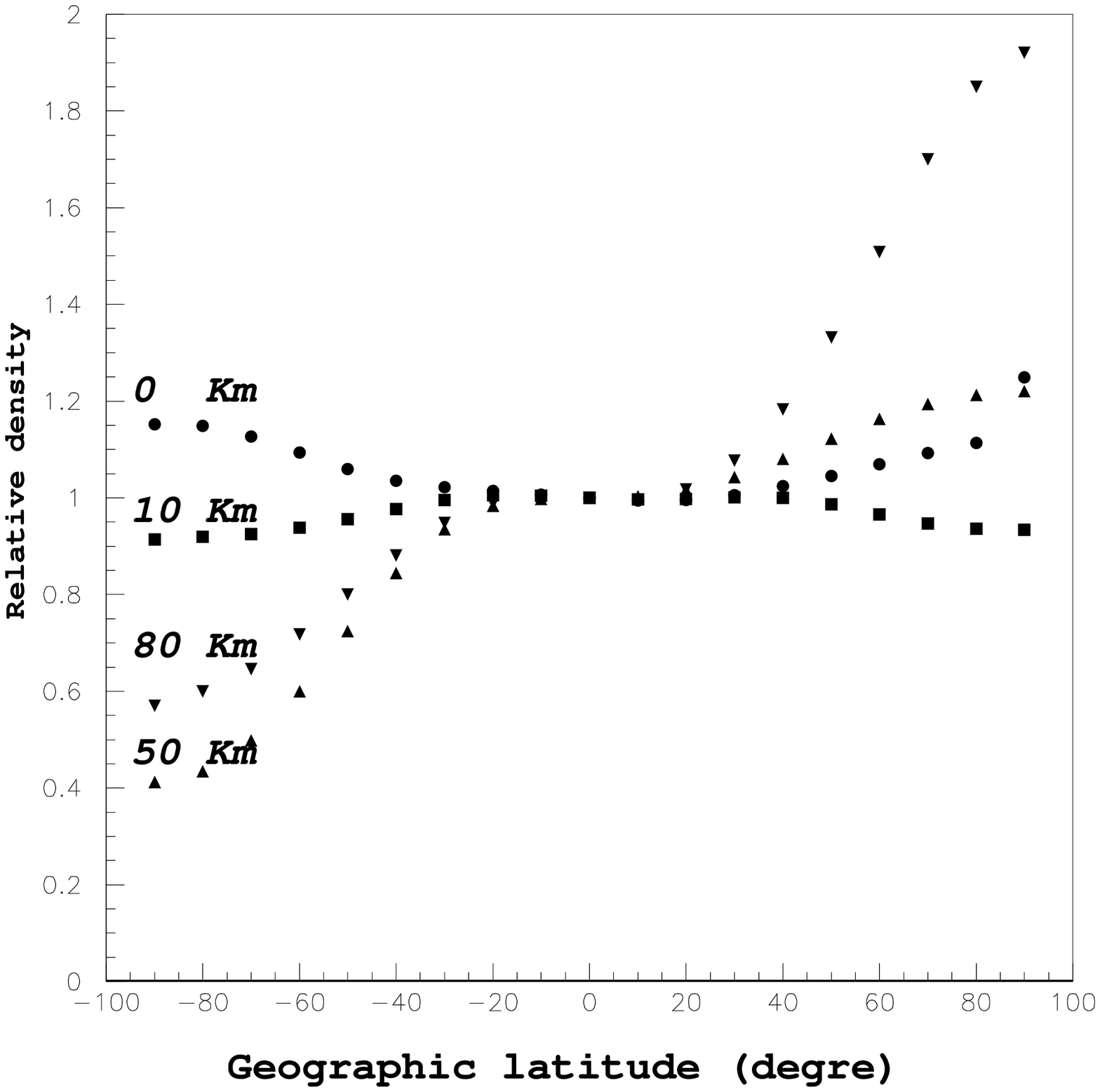,width=16cm}}
\end{center}
\caption{\small Variation of the air density with geographic latitude, for
different altitudes. If the variation is limited to 30 \% for low altitudes, it
can reaches upto a factor 3 for the higher layers( data from \cite{atmos}).
}
\label{fg:densite}
\end{fighere}

%fig 12
\newpage
\begin{fighere}
\begin{center}
 \mbox{\epsfig{file=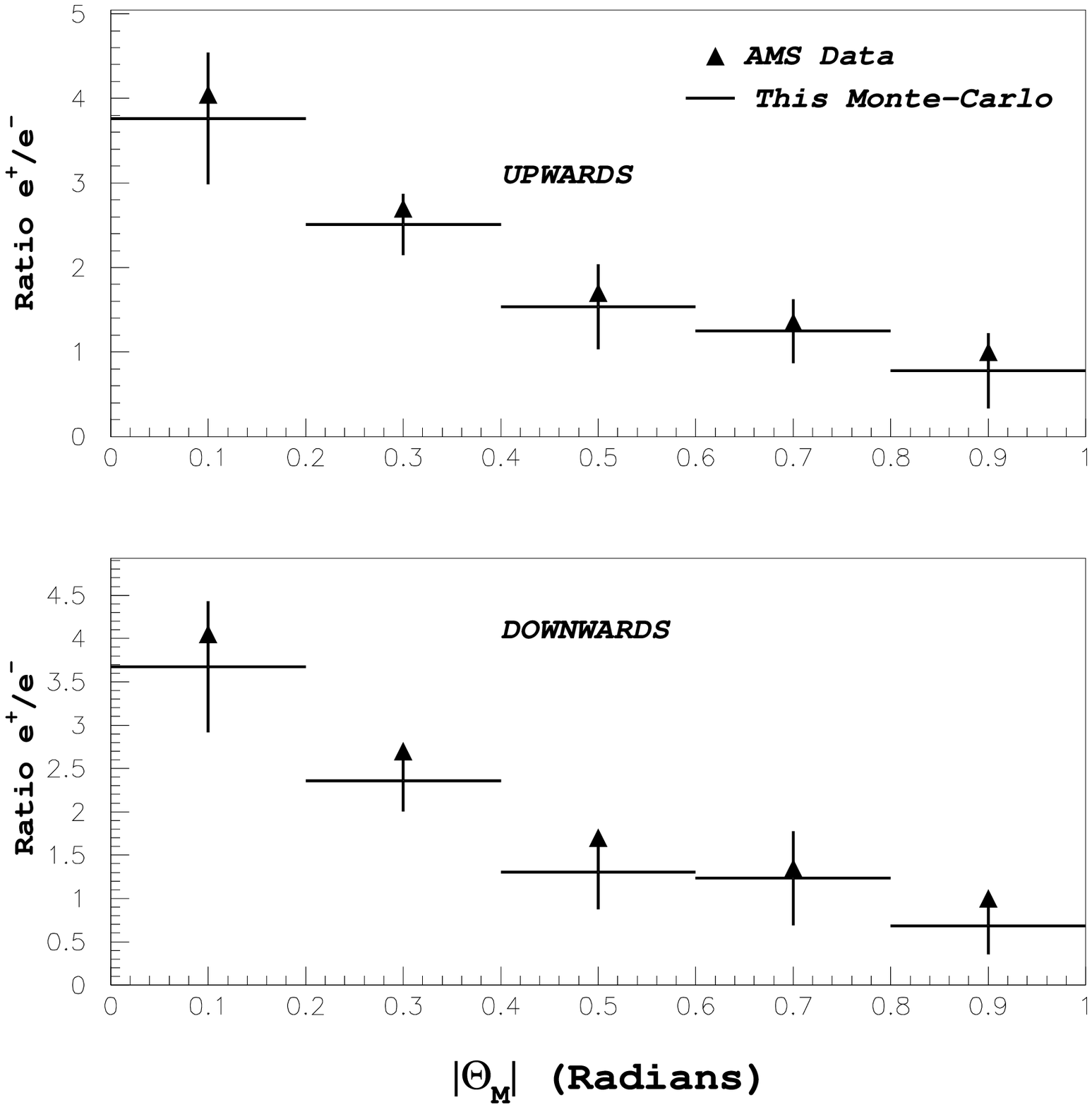,width=16cm}}
\end{center}
\caption{\small Evolution with geomagnetic latitude of the positron-electron yields ratio for upwards and downwards incoming directions.
}
\label{fg:raelec}
\end{fighere}

%fig 13
\newpage
\begin{fighere}
\begin{center}
 \mbox{\epsfig{file=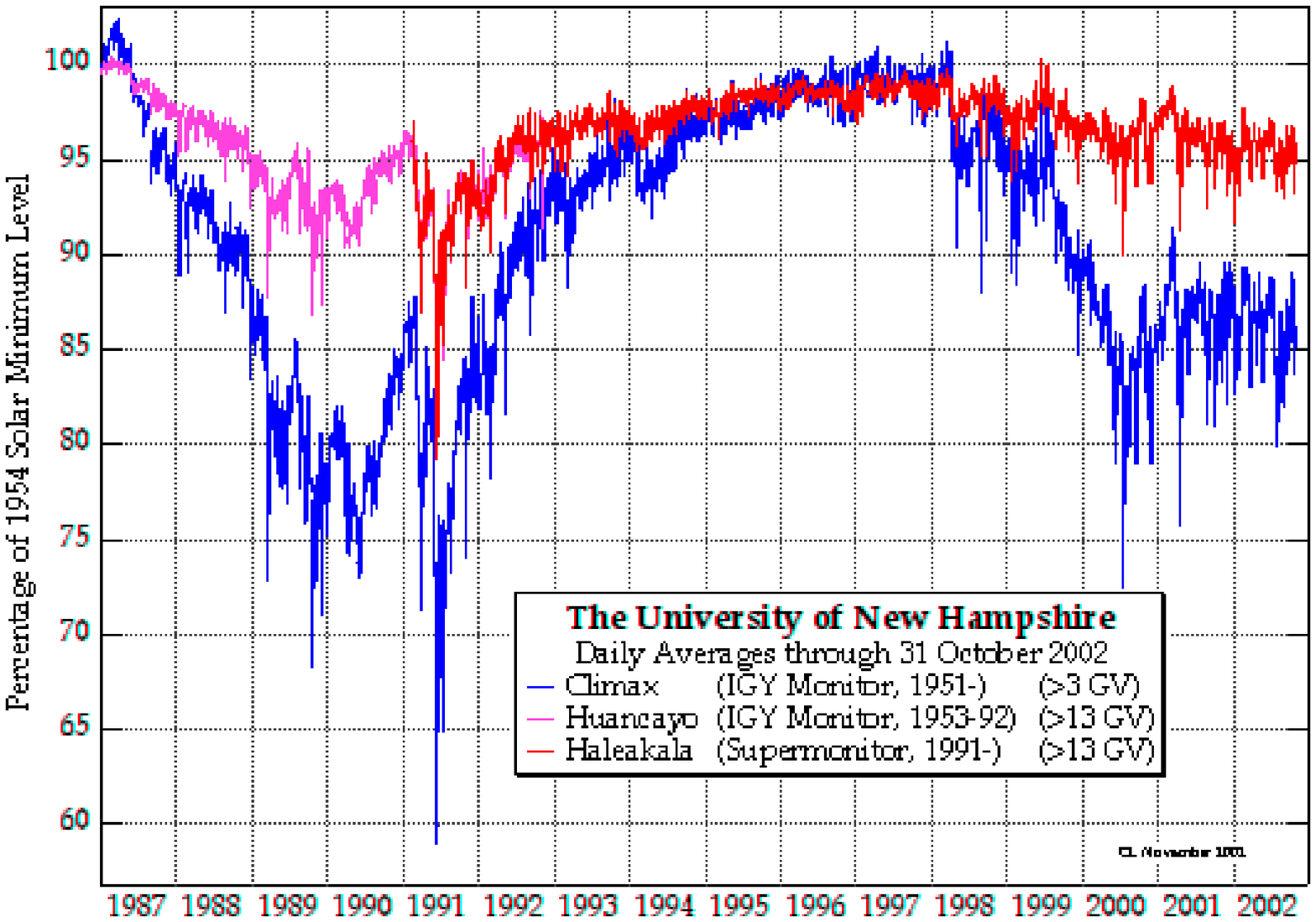,width=16cm}}
\end{center}
\caption{\small Solar activity as measured by the CLIMAX neutron monitor; one
can see on the lower curve the small difference between June 1998 (AMS01) and
July 1994 (CAPRICE) periods.
}
\label{fg:clim}
\end{fighere}

%fig 14
\newpage
\begin{fighere}
\begin{center}
 \mbox{\epsfig{file=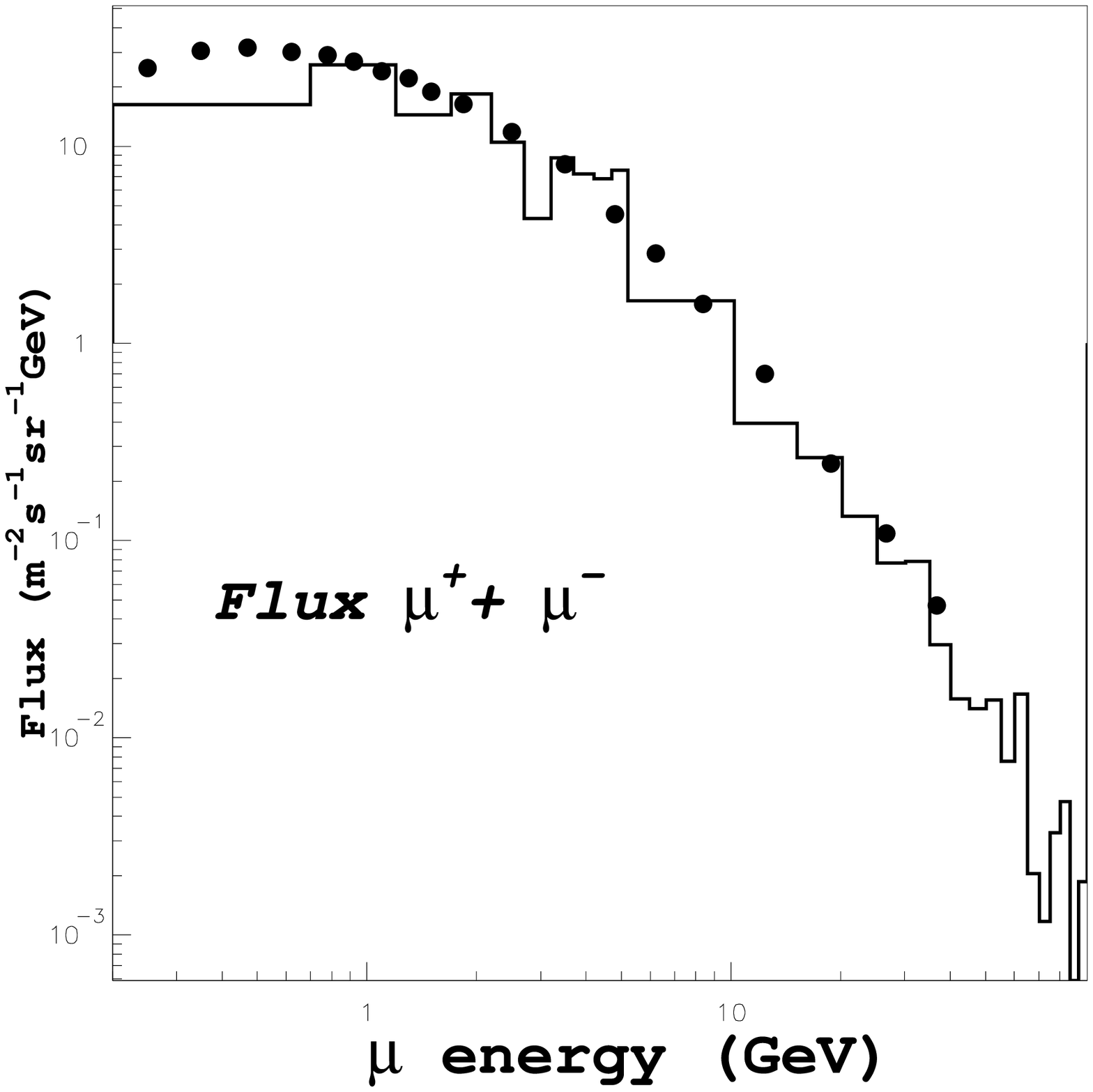,width=16cm}}
\end{center}
\caption{\small Muon spectrum of both charges measured at Lynn Lake by
CAPRICE \cite{caprice}( black circles); the histogram is  the present simulation.
}
\label{fg:muon}
\end{fighere}

%fig 15
\newpage
\begin{fighere}
\begin{center}
 \mbox{\epsfig{file=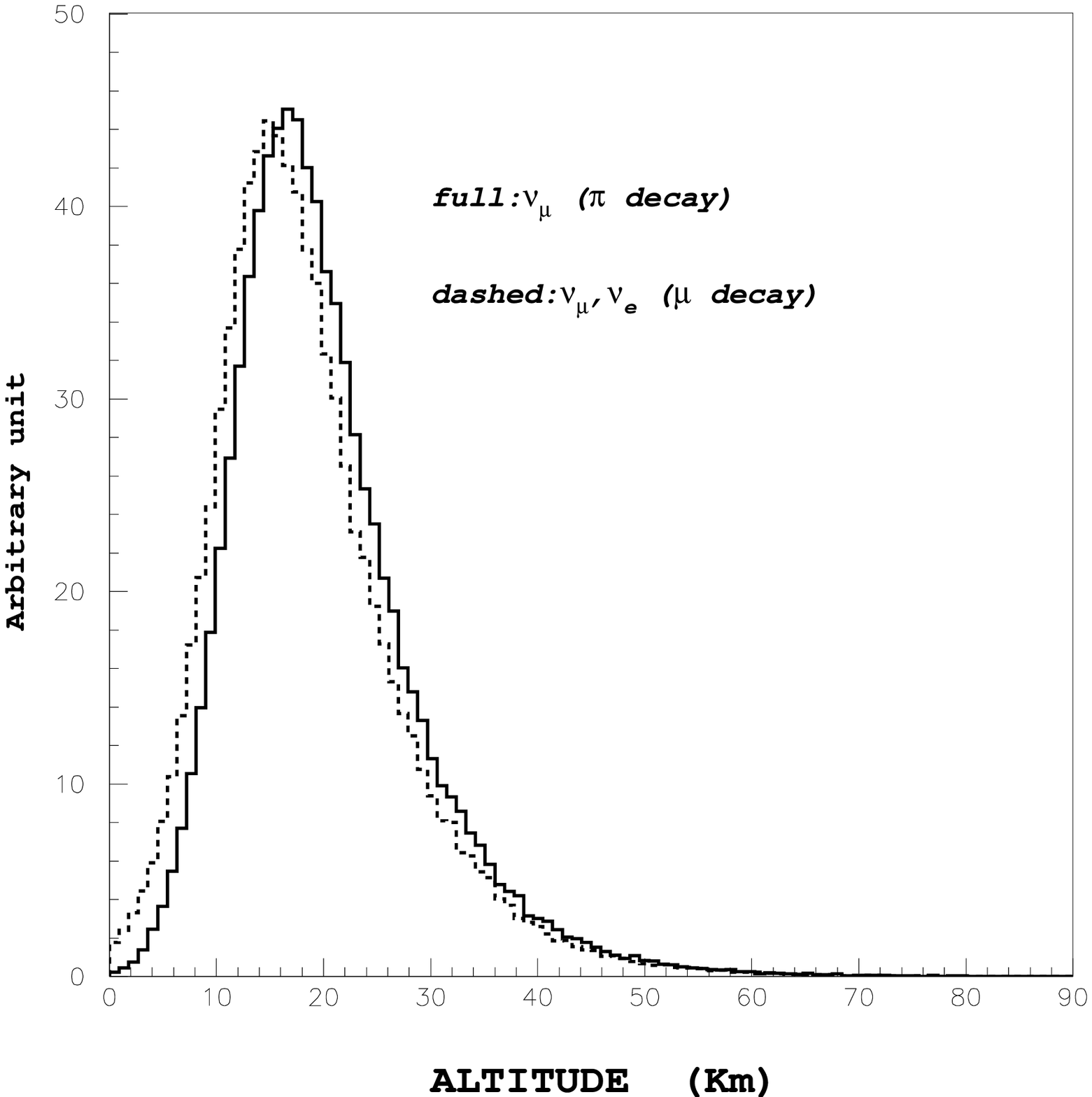,width=16cm}}
\end{center}
\caption{\small Production altitude of neutrinos issued from pions decays and
muons
 decays 
}
\label{fg:altineut}
\end{fighere}
%fig 16
\newpage
\begin{fighere}
\begin{center}
 \mbox{\epsfig{file=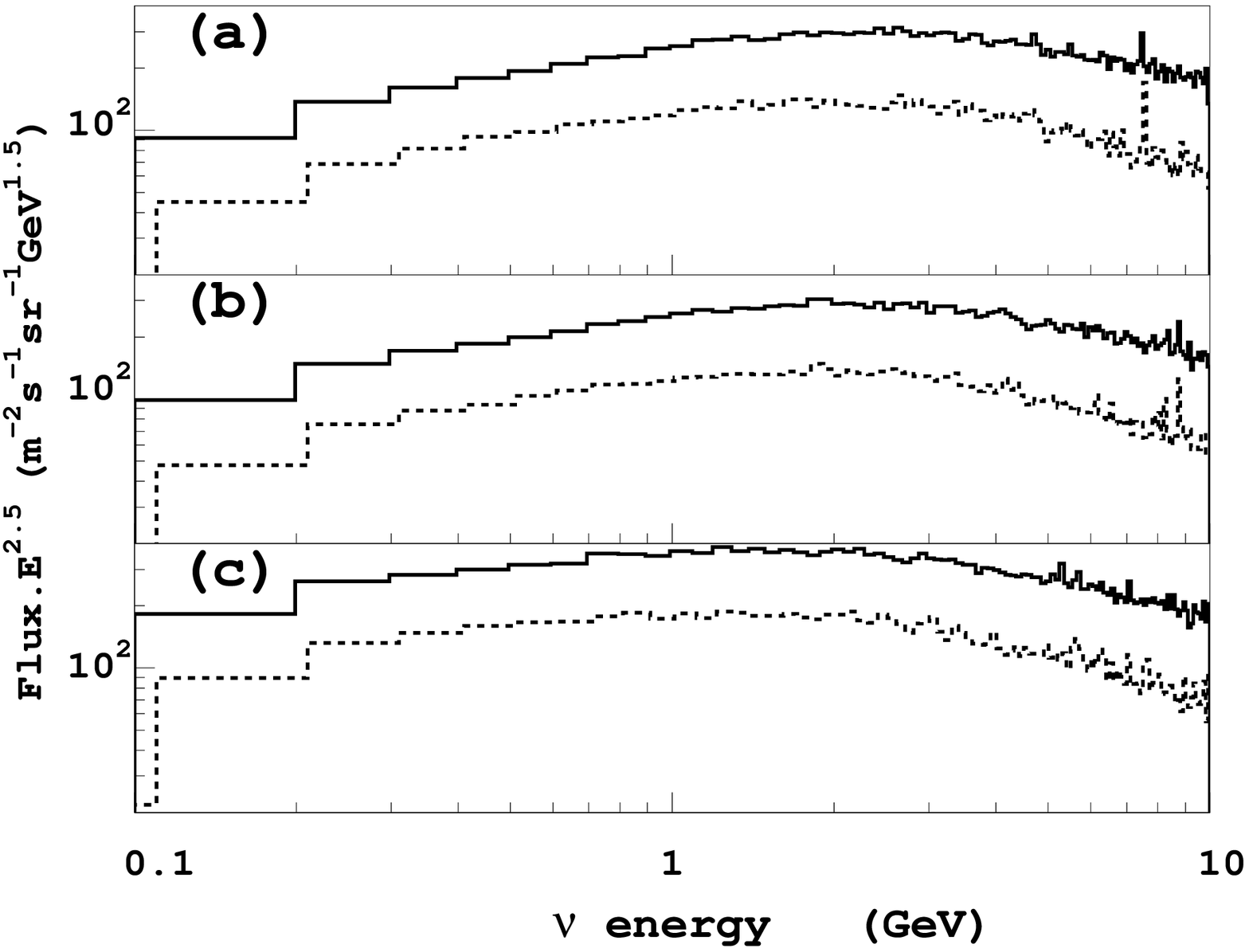,width=16cm}}
\end{center}
\caption{\small Neutrino fluxes averaged on angles for different geomagnetical
latitudes: (a) 0.-0.2 rad, (b) 0.2-0.6 rad, (c) 0.6-1. rad. Full line histogram:
($\nu_{\mu}+{\bar\nu_{\mu}}$). Dashed histogram:($\nu_{e}+{\bar\nu_{e}}$) 
}
\label{fg:enlat}
\end{fighere}

%fig 17
\newpage
\begin{fighere}
\begin{center}
 \mbox{\epsfig{file=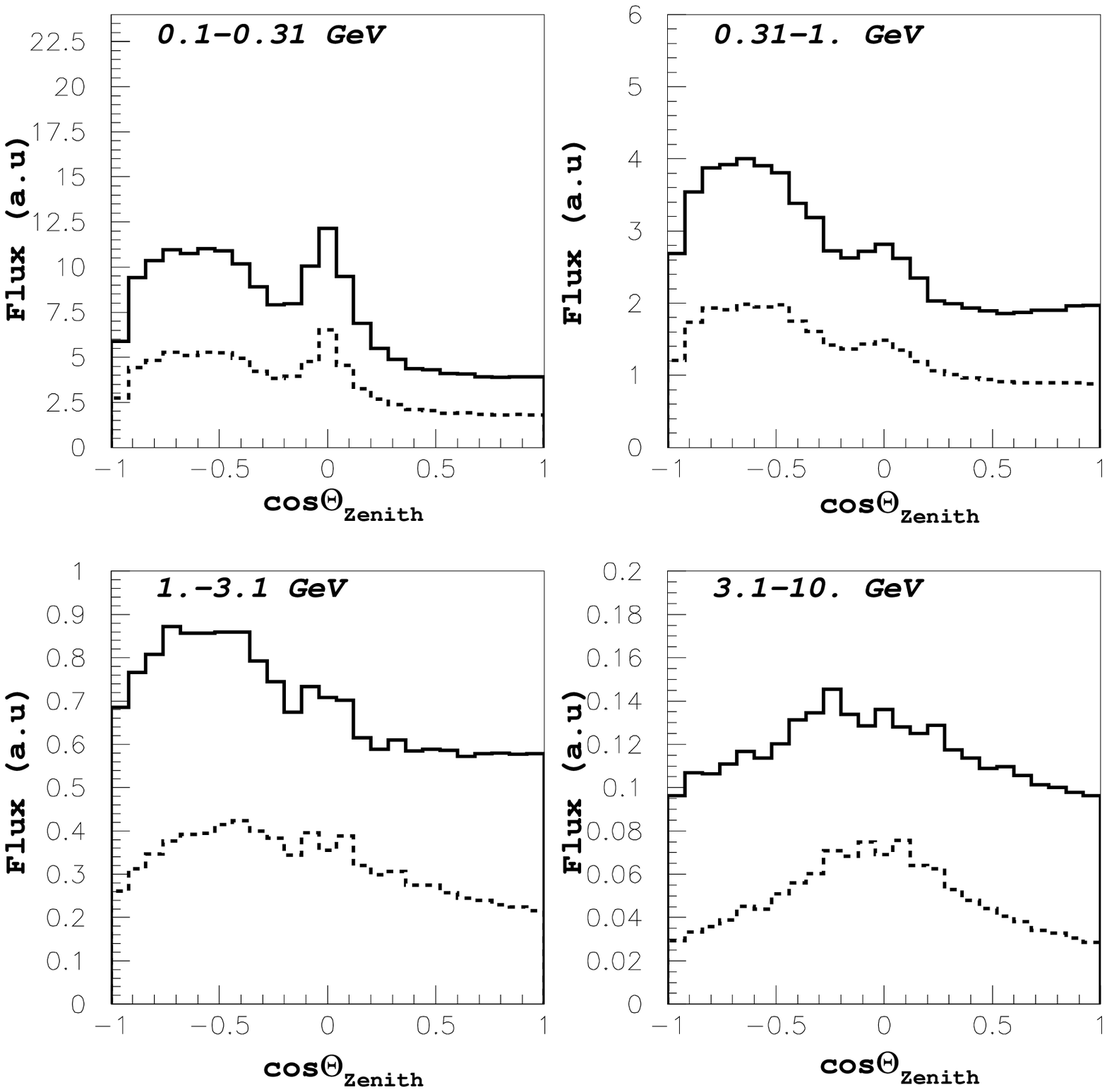,width=16cm}}
\end{center}
\caption{\small In the region of geomagnetical latitude $|\Theta_{M}|\leq$ 0.2 rad, zenithal
angle distributions for 4 different neutrino momentum slices. Full line histogram:
($\nu_{\mu}+{\bar\nu_{\mu}}$). Dashed histogram:($\nu_{e}+{\bar\nu_{e}}$)  
}
\label{fg:zenith1}
\end{fighere}
%\ref{fg:amspro} 
%fig 18
\newpage
\begin{fighere}
\begin{center}
 \mbox{\epsfig{file=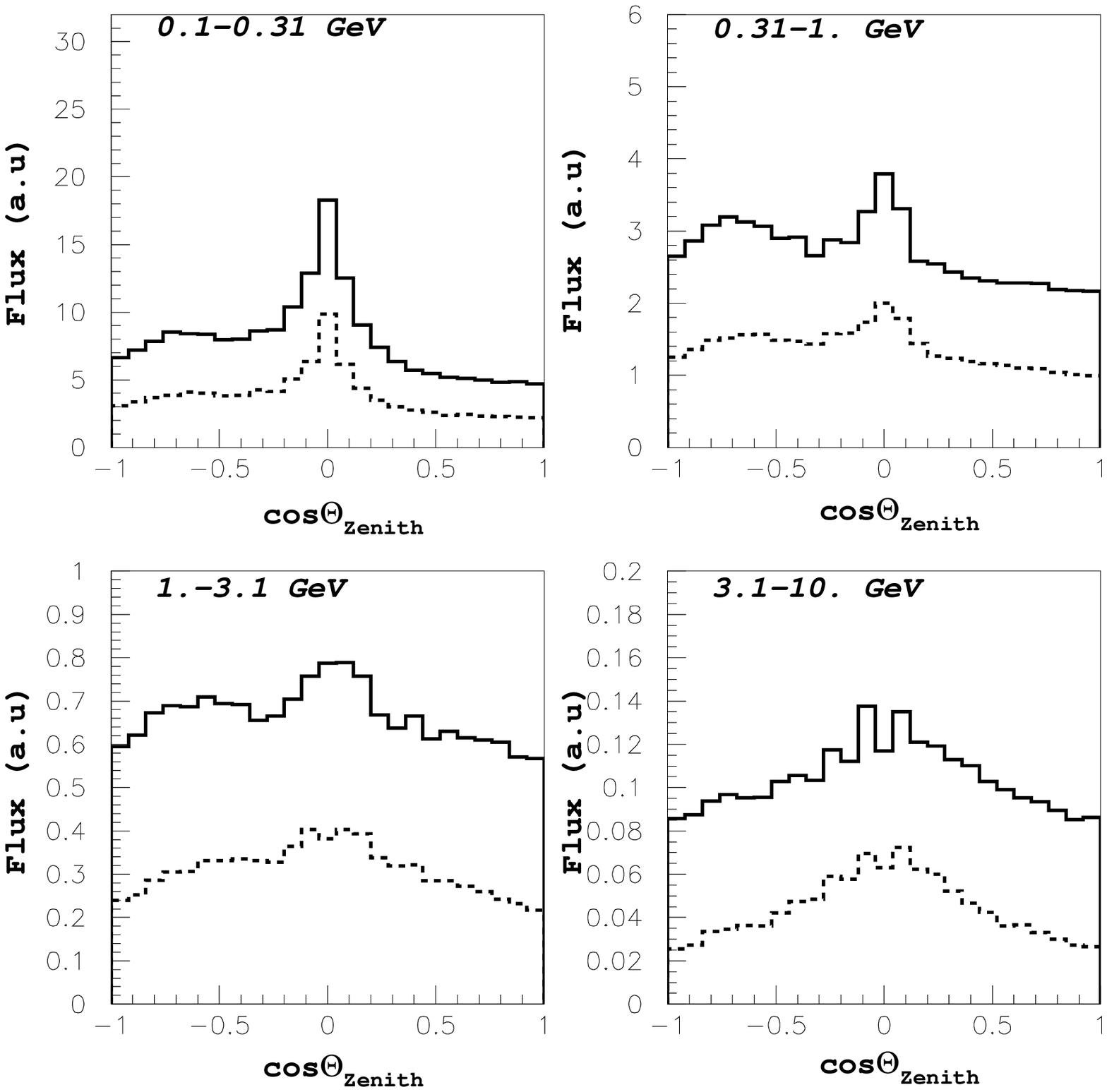,width=16cm}}
\end{center}
%\ref{fg:zenith1}
\caption{\small Same as fig.\ref{fg:zenith1} for the region of geomagnetical latitude 0.2 rad$\leq|\Theta_{M}|
\leq$ 0.6 rad.  
}
\label{fg:zenith2}
\end{fighere}

%fig 19
\newpage
\begin{fighere}
\begin{center}
 \mbox{\epsfig{file=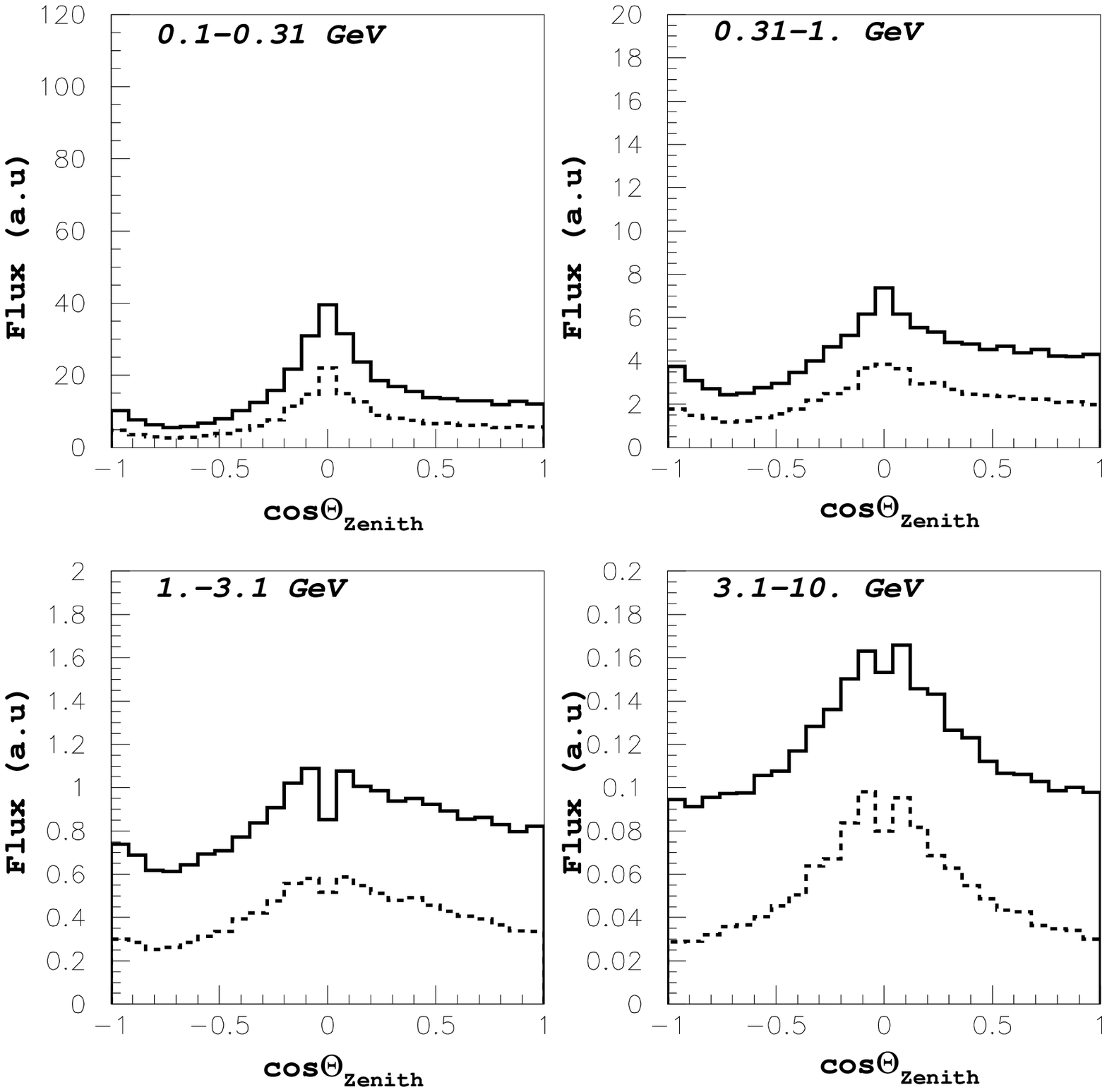,width=16cm}}
\end{center}
\caption{\small Same as fig.\ref{fg:zenith1} for the region of geomagnetical latitude  0.6 rad $\leq|\Theta_{M}|
\leq$ 1. rad. 
}
\label{fg:zenith3}
\end{fighere}

%fig 20
\newpage
\begin{fighere}
\begin{center}
\mbox{\epsfig{file=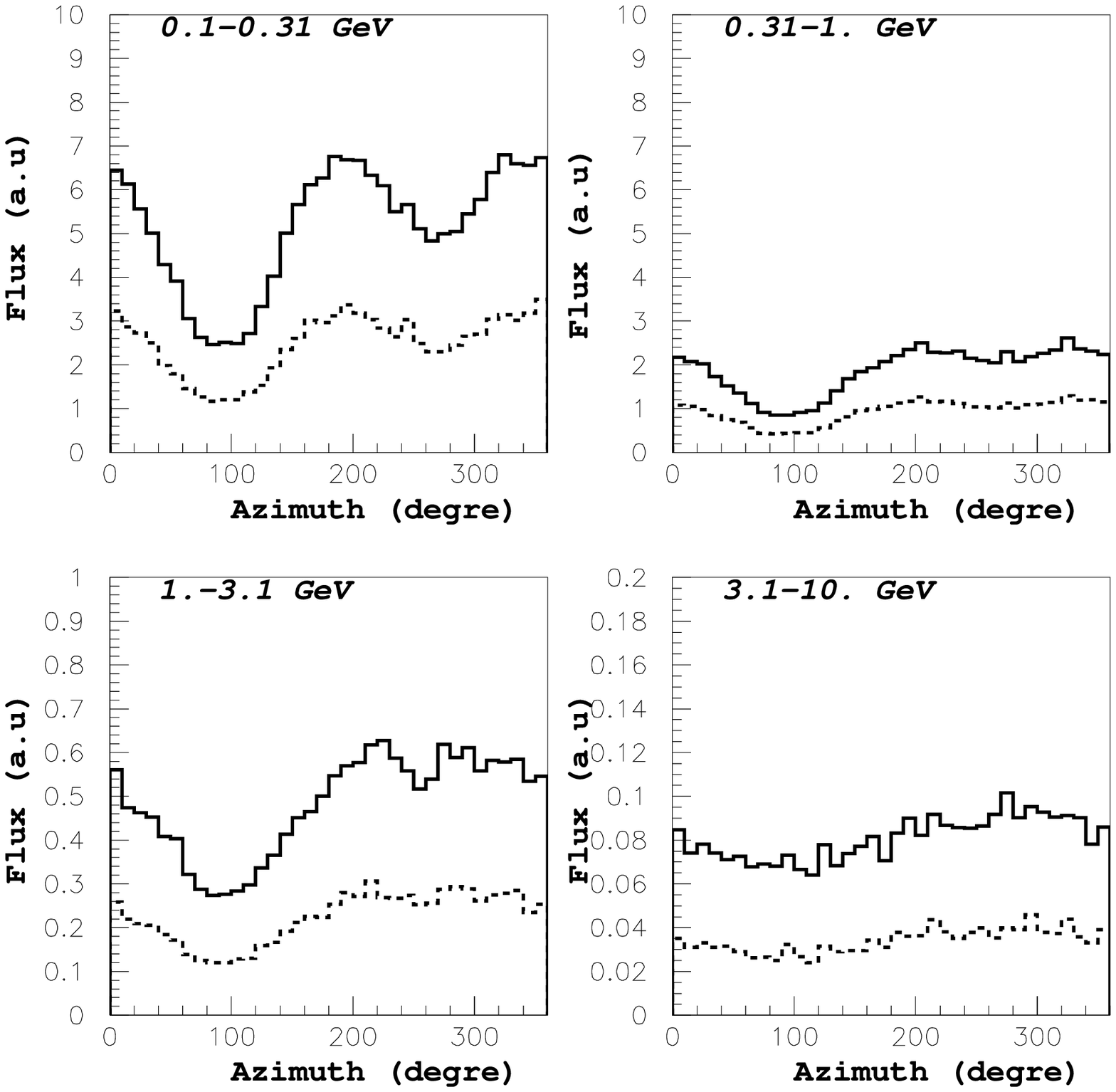,width=16cm}}
\end{center}
\caption{\small In the region of geomagnetical latitude $|\Theta_{M}|\leq$ 0.2
rad., azimuthal angle distributions for 4 different neutrino momentum slices. Full line histogram:
($\nu_{\mu}+{\bar\nu_{\mu}}$). Dashed histogram:($\nu_{e}+{\bar\nu_{e}}$)   
}   
\label{fg:azi1}
\end{fighere}

%fig 21
\newpage
\begin{fighere}
\begin{center}
\mbox{\epsfig{file=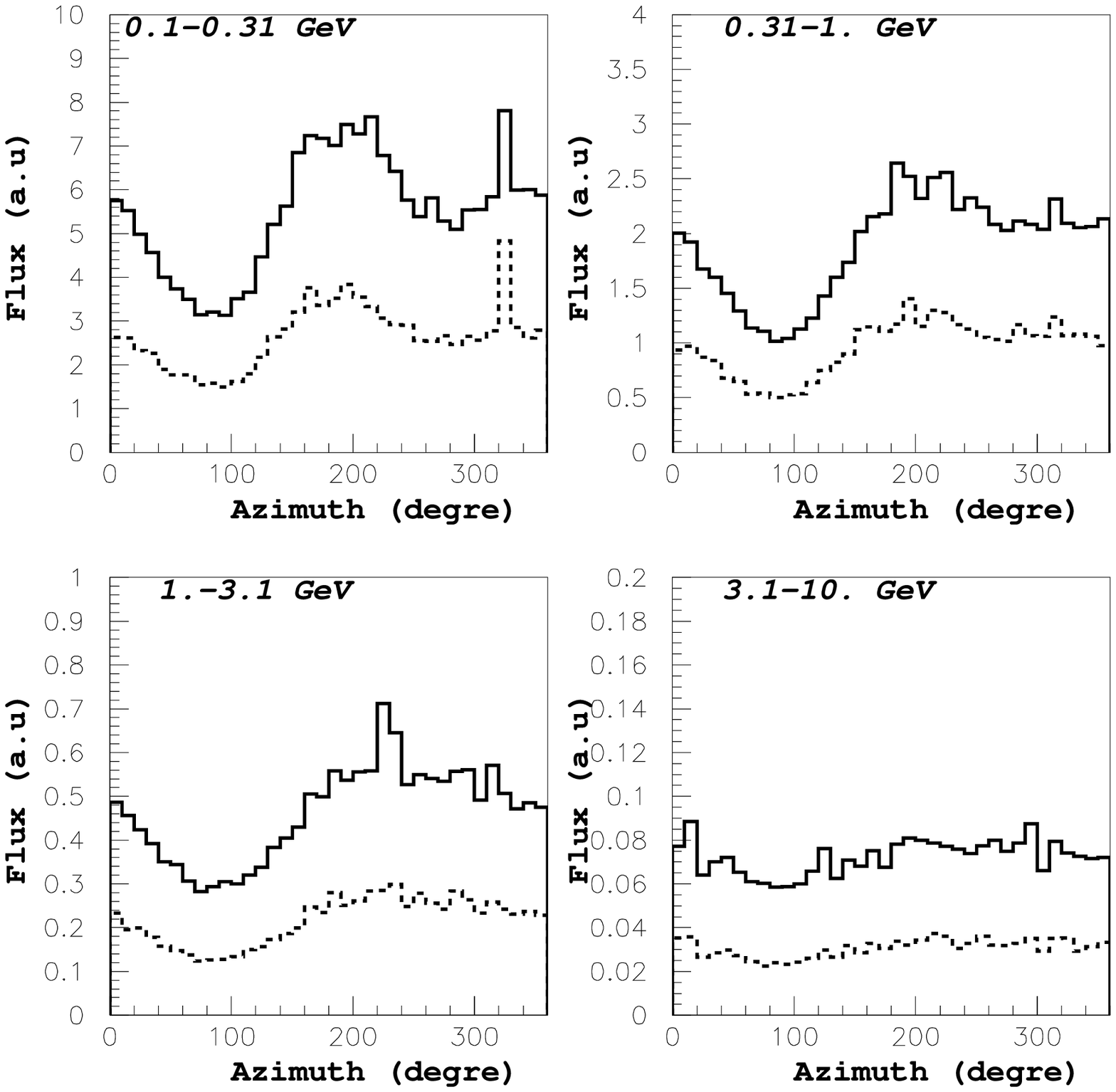,width=16cm}}
\end{center}
%\ref{fg:azi1}
\caption{\small Same as fig.\ref{fg:azi1}  in the region of geomagnetical latitude  0.2 rad$\leq|\Theta_{M}|
\leq$ 0.6 rad. 
}
\label{fg:azi2}
\end{fighere}
%fig 22
\newpage
\begin{fighere}
\begin{center}
 \mbox{\epsfig{file=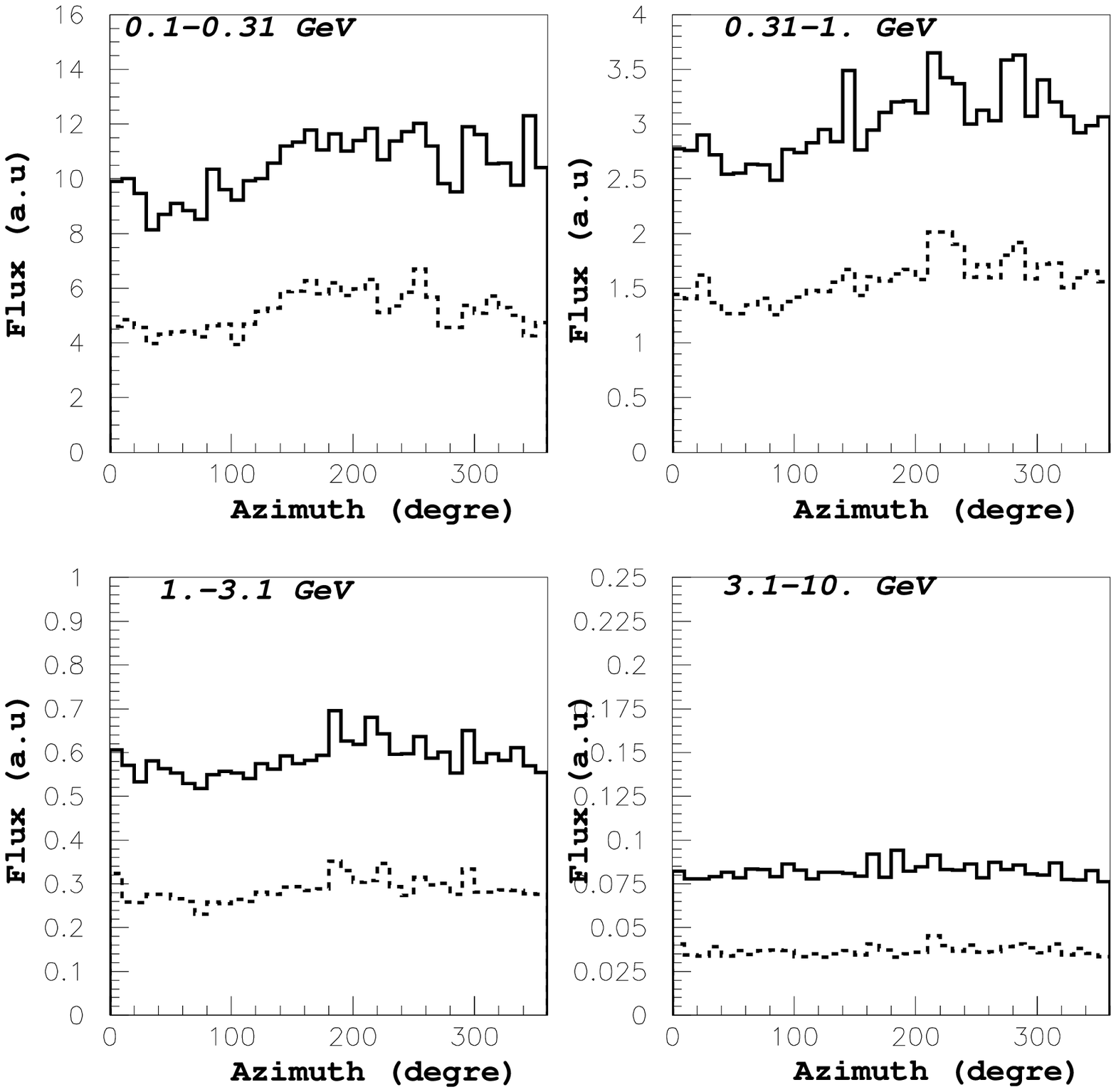,width=16cm}}
\end{center}
\caption{\small Same as fig.\ref{fg:azi1}  in the region of geomagnetical latitude 0.6 rad $\leq|\Theta_{M}|
\leq$ 1. rad. 
}

\label{fg:azi3}
\end{fighere}

%fig 23
\newpage
\begin{fighere}
\begin{center}
 \mbox{\epsfig{file=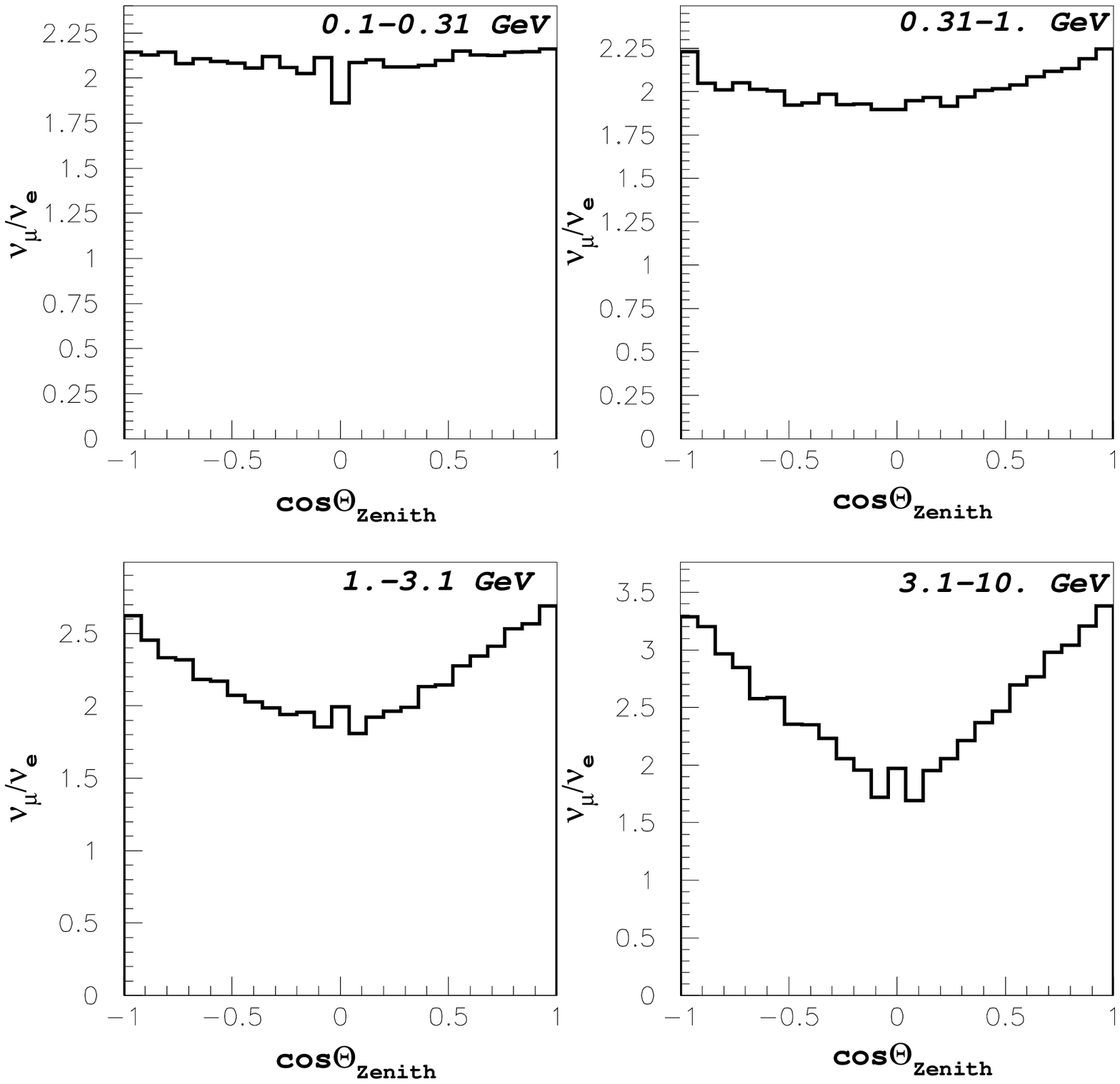,width=16cm}}
\end{center}
\caption{\small In the region of geomagnetical latitude $|\Theta_{M}|\leq$ 0.2 rad,
ratio of ($\nu_{\mu}+{\bar\nu_{\mu}}$) flux over ($\nu_{e}+{\bar\nu_{e}}$) flux
distributions for 4 different neutrino momentum slices.  
}
\label{fg:razen1}
\end{fighere}
%fig 24
\newpage
\begin{fighere}
\begin{center}
 \mbox{\epsfig{file=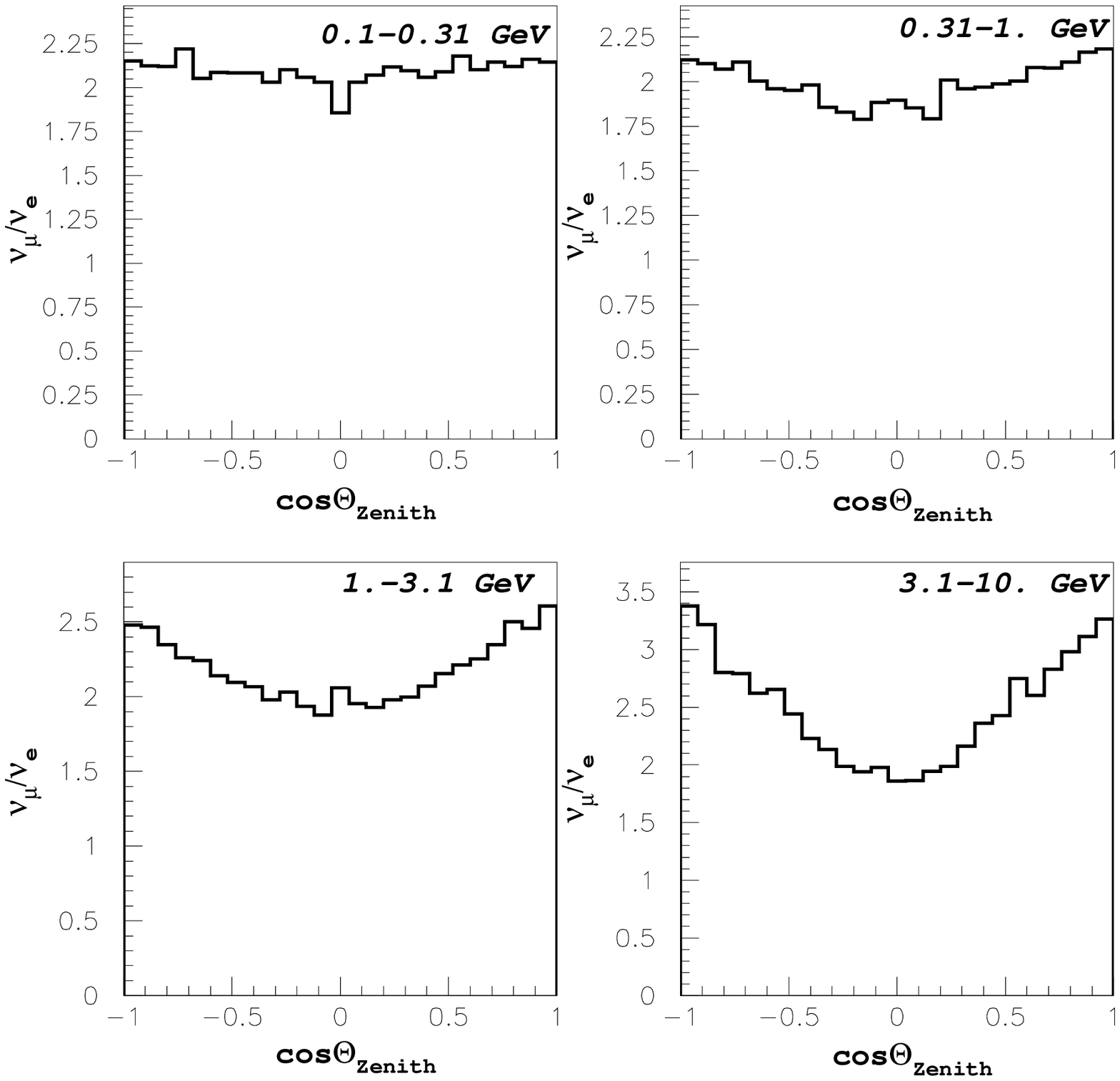,width=16cm}}
\end{center}
%\ref{fg:razen1}
\caption{\small Same as fig.\ref{fg:razen1} in the region of geomagnetical latitude  0.2 rad$\leq|\Theta_{M}|
\leq$ 0.6 rad.
}

\label{fg:razen2}
\end{fighere}
%fig 25
\newpage
\begin{fighere}
\begin{center}
 \mbox{\epsfig{file=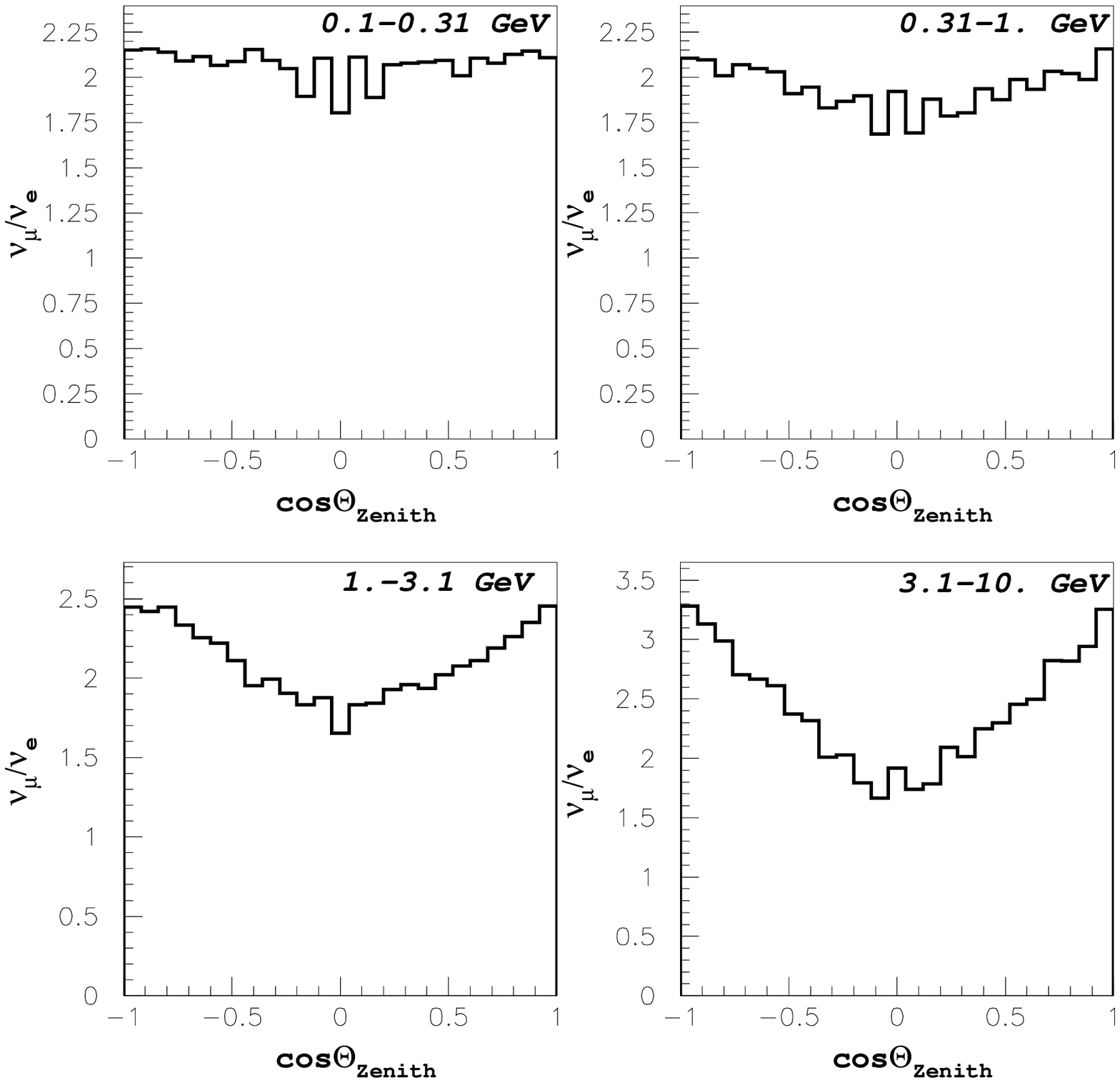,width=16cm}}
\end{center}
\caption{\small Same as fig.\ref{fg:razen1} in the region of geomagnetical latitude 0.6 rad $\leq|\Theta_{M}|
\leq$ 1. rad. 
}
\label{fg:razen3}
\end{fighere}

%fig 26
\newpage
\begin{fighere}
\begin{center}
 \mbox{\epsfig{file=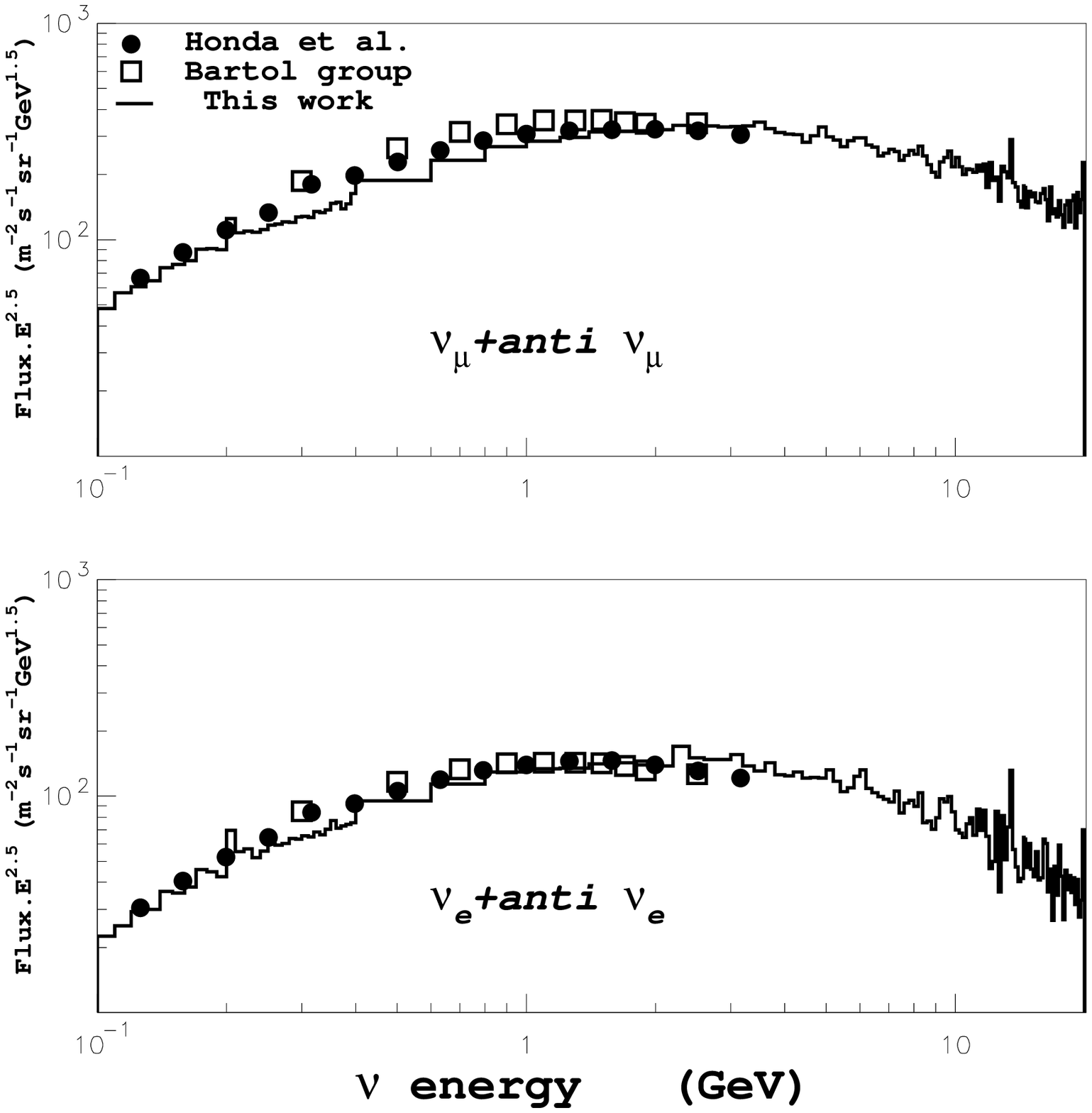,width=16cm}}
\end{center}
\caption{\small Flux$\times E_\nu^{2.5}$ and averaged on angles as a function of neutrino energy.
 On purpose of comparison, we used here
the same primary spectra as Honda et al.\cite{honda}; black dots are M.C from
ref.\cite{honda}, squares from ref.\cite{gaiss} at solar maximum.   
}

\label{fg:hondspec}
\end{fighere}

%fig 27
\newpage
\begin{fighere}
\begin{center}
 \mbox{\epsfig{file=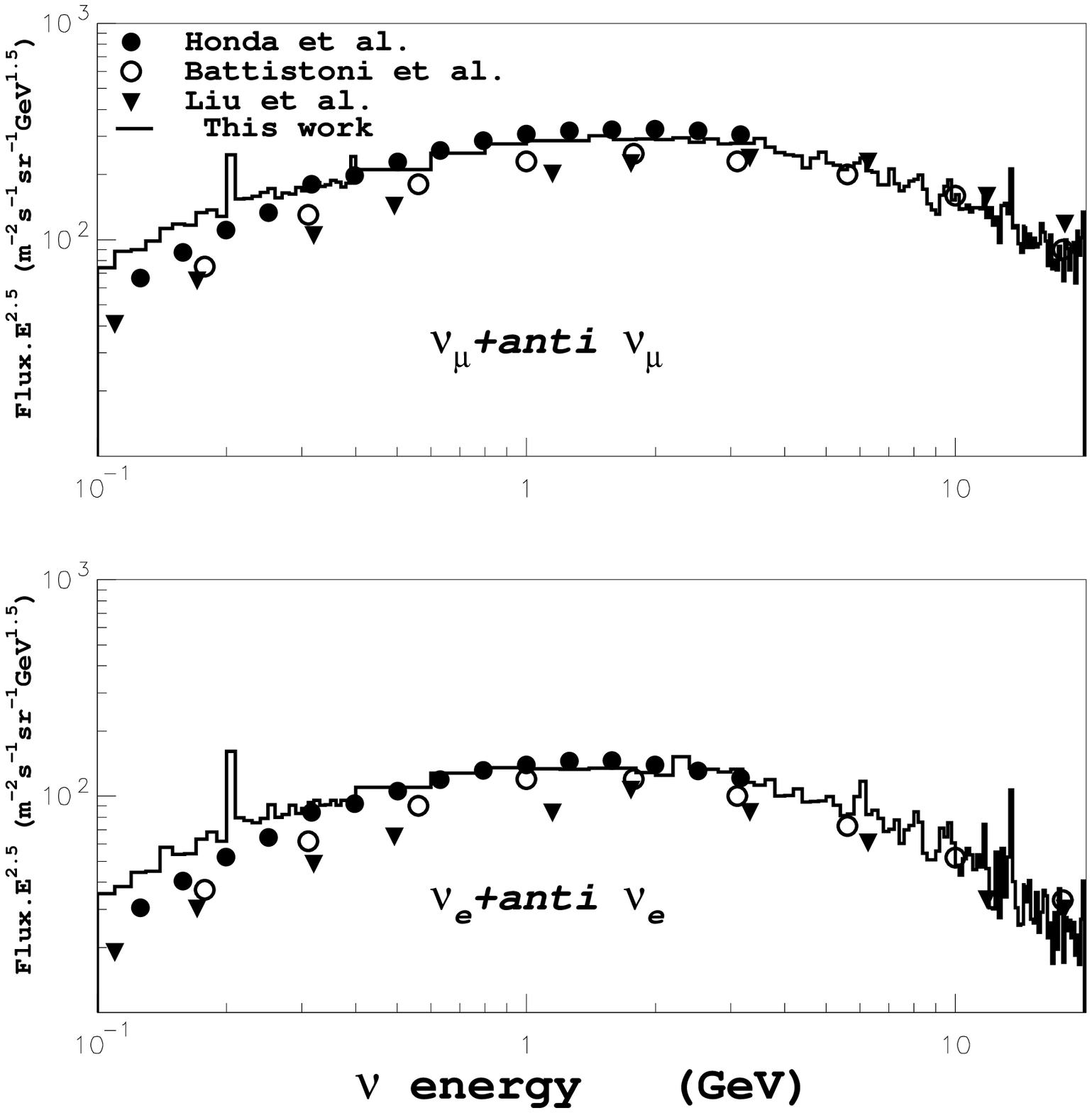,width=16cm}}
\end{center}
\caption{\small Flux$\times E_\nu^{2.5}$ and averaged on angles as a function of neutrino energy. Here
we used our AMS primary spectrum, corrected for solar modulation corresponding
to years 1996,1997,1998; black dots are M.C from ref.\cite{honda}, circles from
ref.\cite{batti} (solar maximum), triangles from ref.\cite{buen3}.   
}
\label{fg:ninospec}
\end{fighere}

%fig 28
\newpage
\begin{fighere}
\begin{center}
 \mbox{\epsfig{file=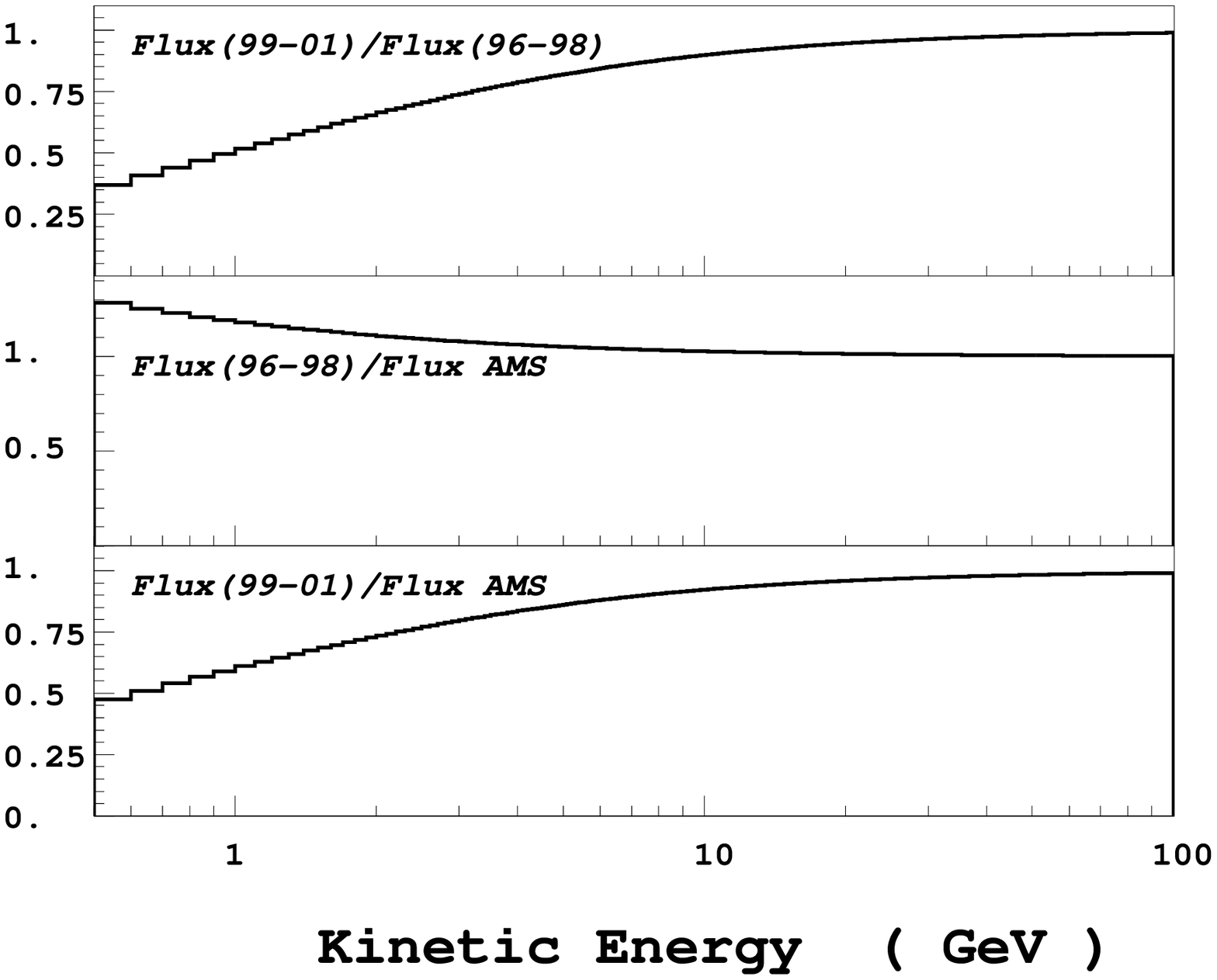,width=16cm}}
\end{center}
\caption{\small Ratio of primary protons spectra for three different solar
activities periods: SK1=1996+1997+1998 ; SK2=1999+2000+2001; AMS: June 1998.
Computed from ref.\cite{honda} formulaes and CLIMAX neutron
monitor data \cite{climax}. 
}
\label{fg:climsk}
\end{fighere}

%fig 29
\newpage
\begin{fighere}
\begin{center}
 \mbox{\epsfig{file=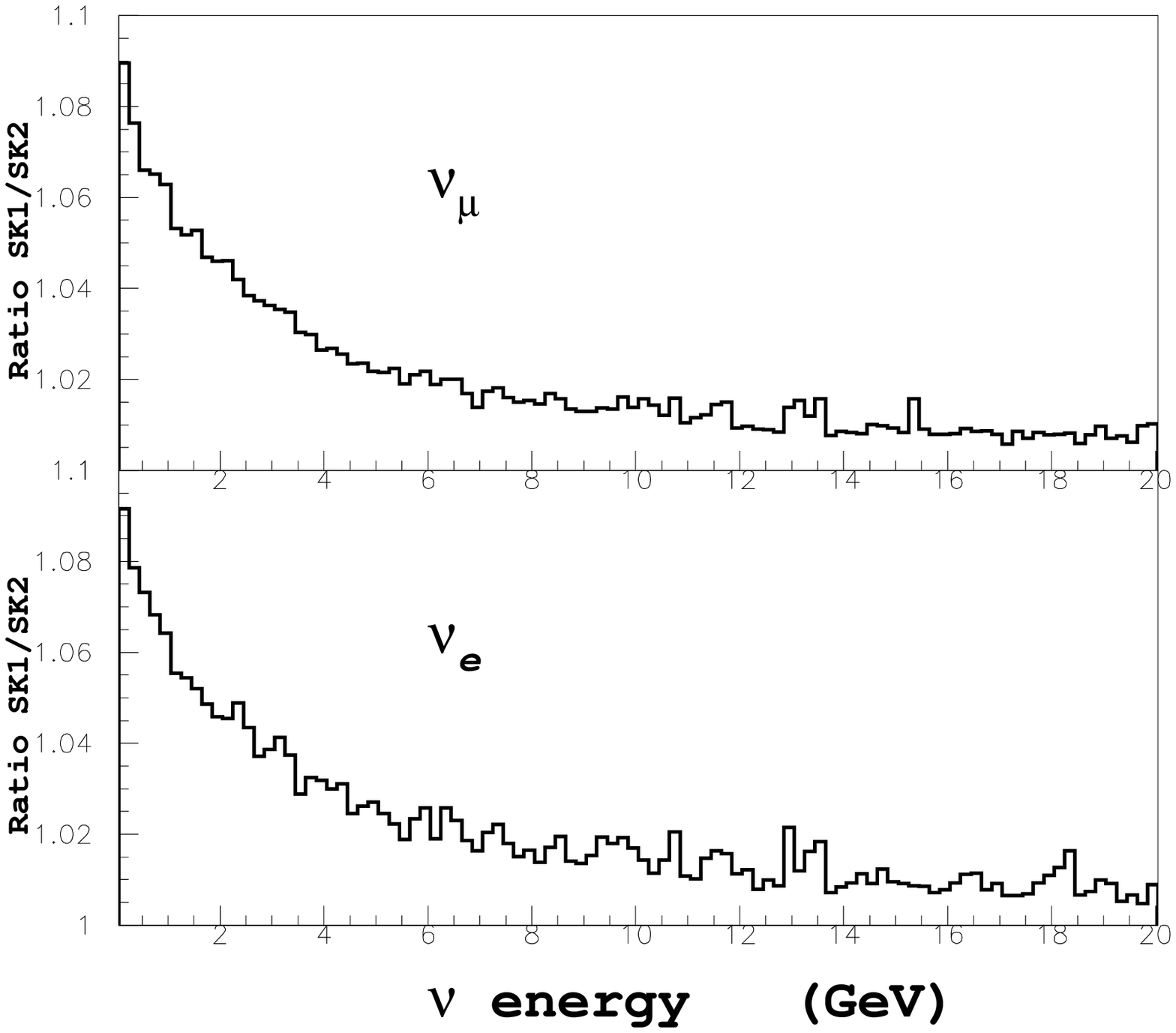,width=16cm}}
\end{center}
\caption{\small Ratio of neutrino spectra between SK1 (years 1996,1997,1998) and
SK2 (years 1999,2000,2001), using the modulation formula of ref.\cite{honda}.
}
\label{fg:ninosk2}
\end{fighere}

%fig 30
\newpage
\begin{fighere}
\begin{center}
 \mbox{\epsfig{file=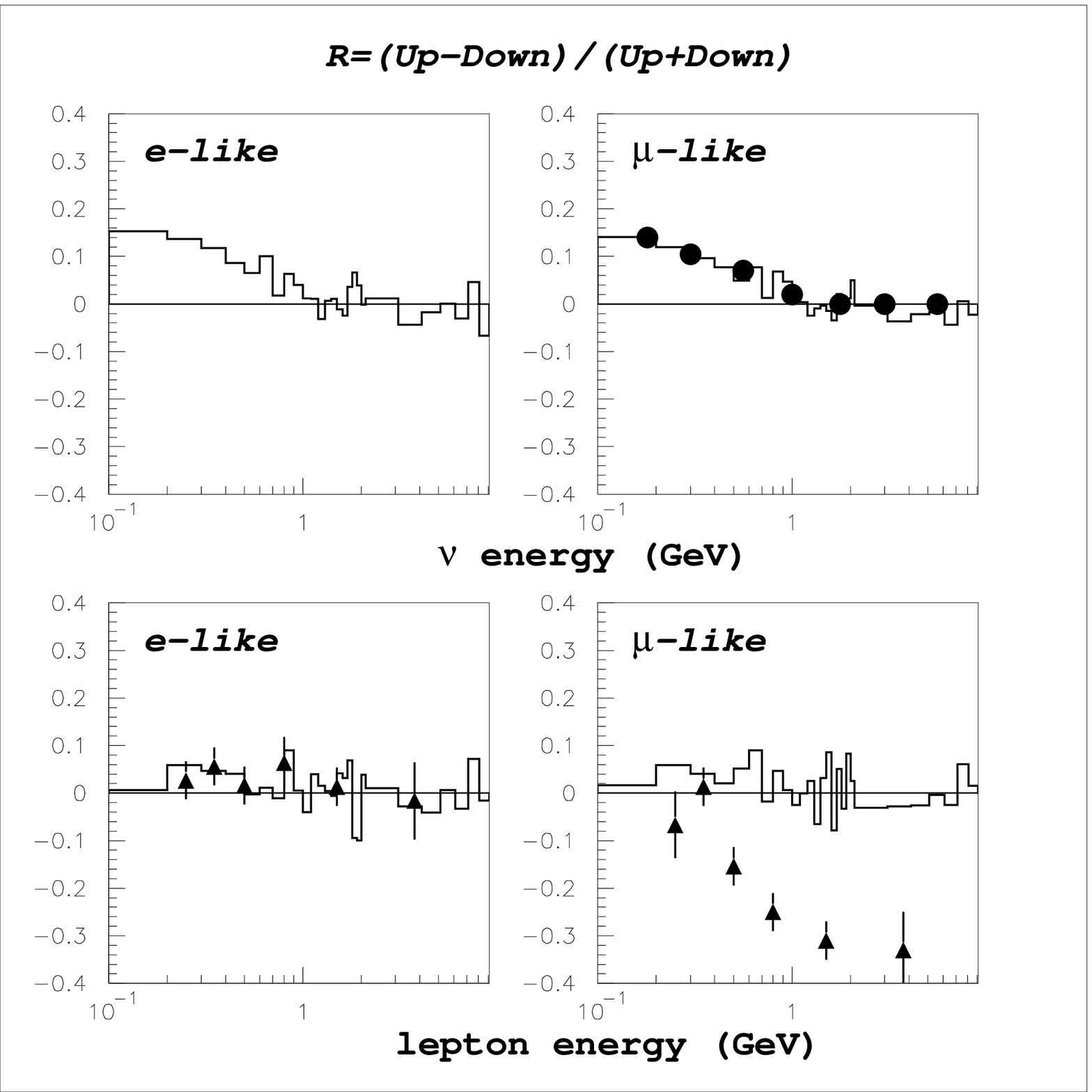,width=16cm}}
\end{center}
\caption{\small The up-down assymetry is defined as the difference of up and
down neutrino fluxes divided by their sum. The two top histograms use the
neutrinos angles and momentum; the bottom ones use smeared induced leptons
corresponding variables.The black dots are the result from a 3-D
Fluka recent version M.C. (ref.\cite{batti2}), and triangles are Super Kamiokande points
from ref.\cite{learned} 
}
\label{fg:updown}
\end{fighere}

%fig 31
\newpage
\begin{fighere}
\begin{center}
 \mbox{\epsfig{file=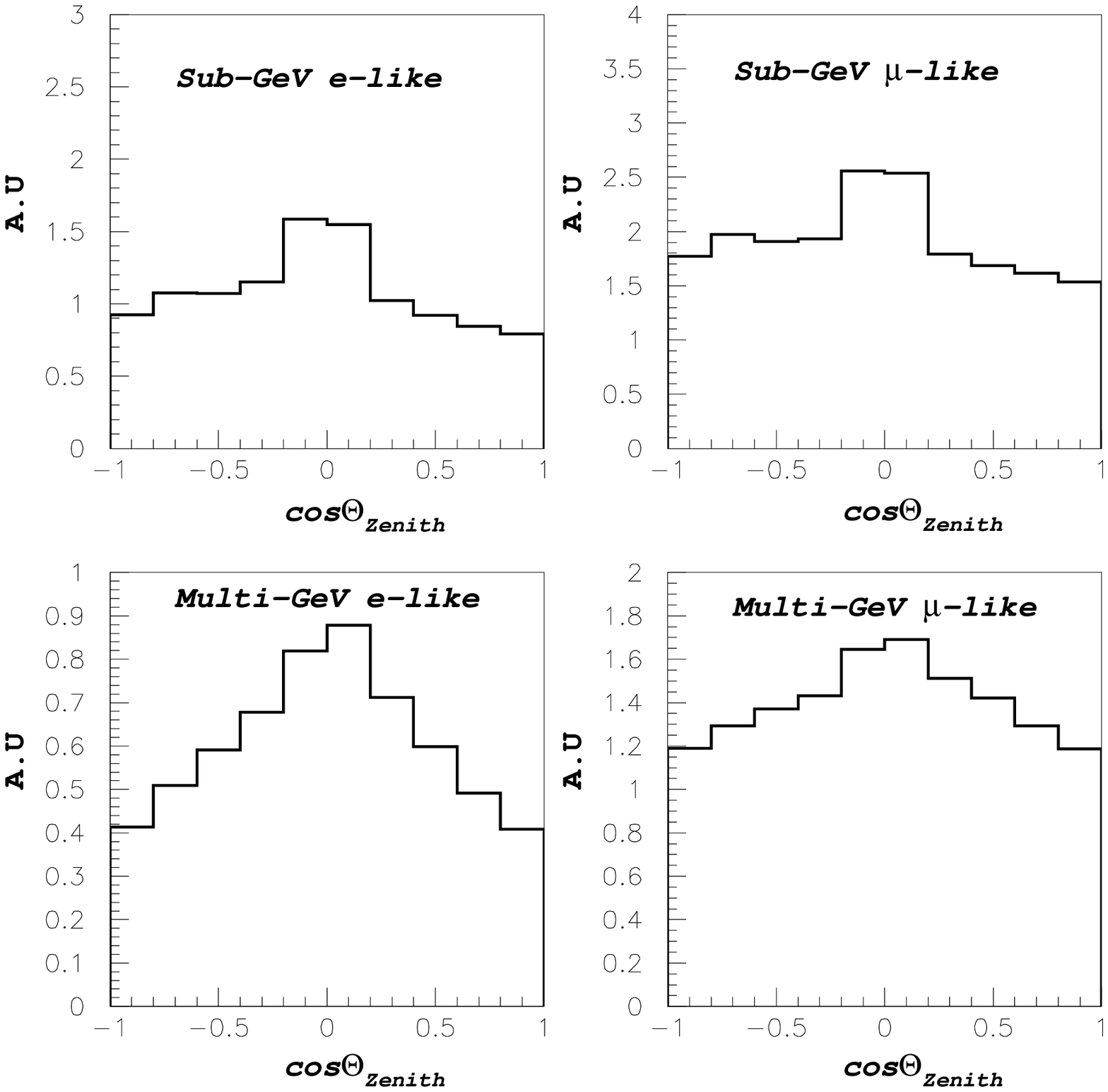,width=16cm}}
\end{center}
\caption{\small Variation with the zenith angle of neutrinos fluxes at
Super-Kamiokande latitude. Here, no smearing is applied to the neutrinos angles
and momentum. 
}
\label{fg:zenit}
\end{fighere}

%fig 32
\newpage
\begin{fighere}
\begin{center}
 \mbox{\epsfig{file=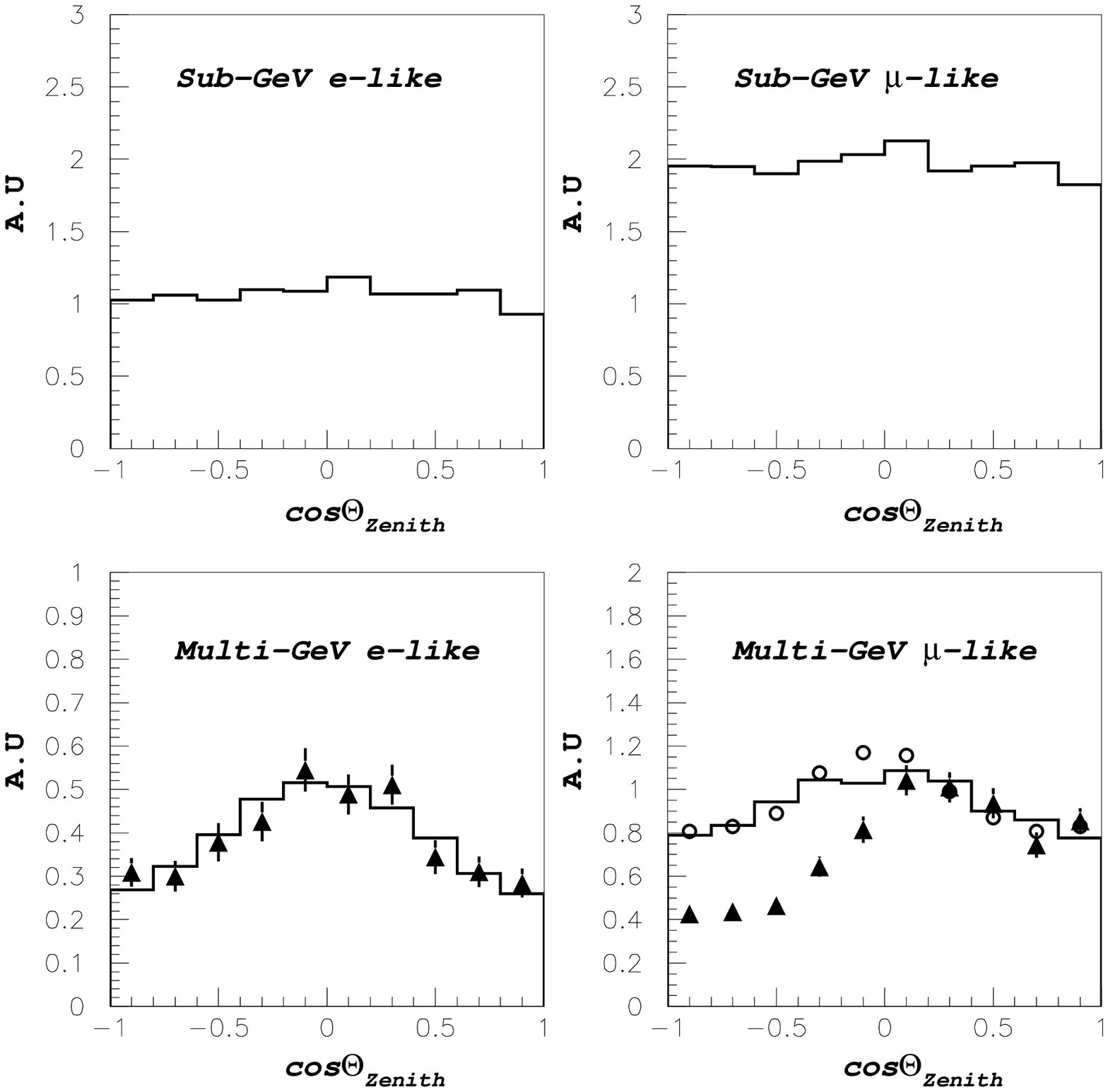,width=16cm}}
\end{center}
\caption{\small Variation with the zenith angle of neutrinos fluxes at
Super-Kamiokande latitude. Smearing is applied to the neutrinos angles
and momentum to reproduce the incertitude on the neutrino variables when the
induced lepton is used.We have reproduced for the multi-GeV views the results of
  S.K \cite{shiozawa}(black triangles), where we normalized at equal surface for the e-like
  histogram, and only at the three last bins in cos$\theta$ for the $\mu-like$
  plot. Open circles are the Honda's M.C. results shown
  in\cite{shiozawa}. 
}

\label{fg:zenitsk}
\end{fighere}

%fig 33
\newpage
\begin{fighere}
\begin{center}
 \mbox{\epsfig{file=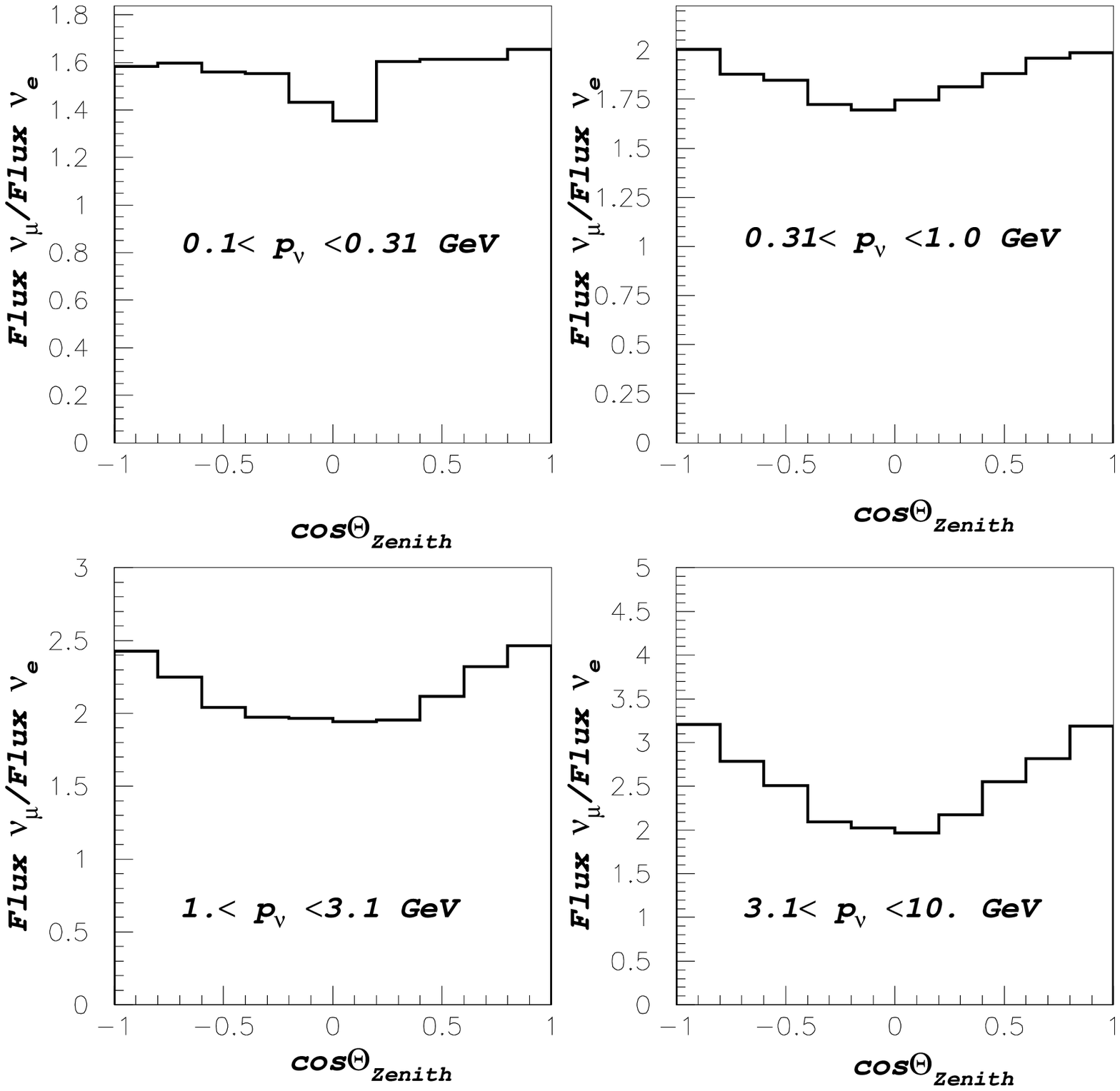,width=16cm}}
\end{center}
\caption{\small The ratio of fluxes ($\nu_{\mu}+\bar\nu_{\mu}$) / ($\nu_{e}+\bar\nu_{e}$) 
for 4 different neutrino momentum slices, at the Super-Kamiokande
latitude, as a function of the zenith angle.  
}
\label{fg:rapsk}
\end{fighere}

%fig 34
\newpage
\begin{fighere}
\begin{center}
\mbox{\epsfig{file=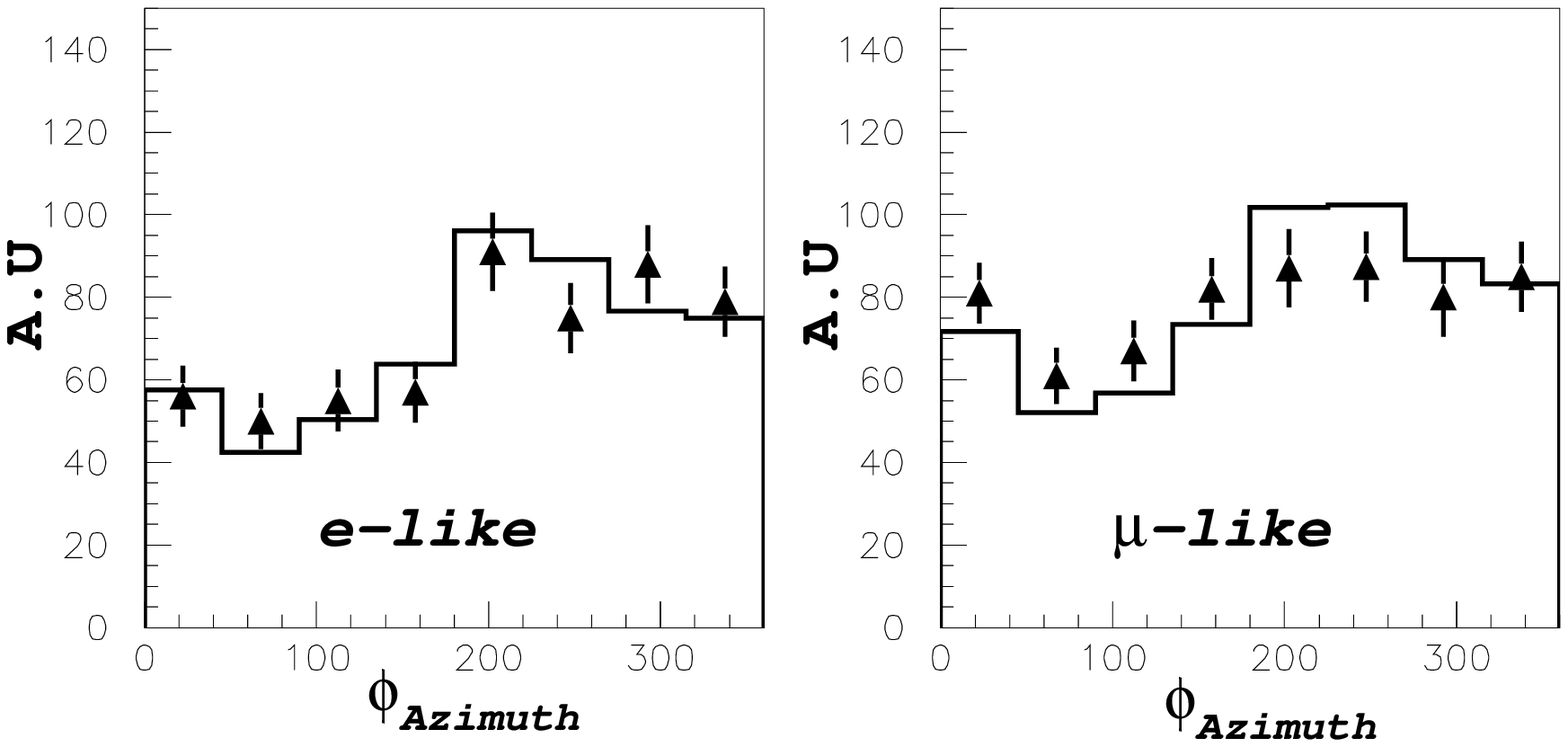,width=16cm}}
\end{center}
\caption{\small Comparison of Super-Kamiokande ( black triangles) and M.C for
azimuthal direction of the incoming neutrino; M.C is smeared with S.K angular
resolution and weighted proportionnal to neutrino cros-section;  normalisation is done to the
total number of S.K events. Events selection is: $\leq0.4$ p $\leq 3.$ GeV, and
abs(cos $\theta_{Zenith})\geq 0.5$ ) No oscillation is present in MC. 
}
\label{fg:estouest}
\end{fighere}

\end{document}